\documentclass[letterpaper,journal]{IEEEtran}
\usepackage{amsmath,amsfonts}
\usepackage{algorithmic}
\usepackage{algorithm}
\usepackage{array}
\usepackage[caption=false,font=normalsize,labelfont=sf,textfont=sf]{subfig}
\usepackage{textcomp}
\usepackage{hyperref}
\usepackage{stfloats}
\usepackage{url}
\usepackage{verbatim}
\usepackage{graphicx}
\usepackage{booktabs}   
\usepackage{cite}

\begin{document}

\title{DIA: A Deep Reinforcement Learning-Based Interest-Aware ABR Scheme for Cloud Gaming}

\author{Zhe Zhang,~\IEEEmembership{Senior Member,~IEEE,} Haiwei Dong,~\IEEEmembership{Senior Member,~IEEE,} Marc St-Hilaire,~\IEEEmembership{Senior Member,~IEEE,} Abdulmotaleb El Saddik,~\IEEEmembership{Fellow,~IEEE}
\thanks{Zhe Zhang is with the School of Communications and Information Engineering, Nanjing University of Posts and Telecommunications, Nanjing, Jiangsu, China (E-mail: zhezhang@njupt.edu.cn).}
\thanks{Haiwei Dong is with Ottawa Research Center, Huawei Technologies Canada, Ottawa, ON K2K 3J1, Canada, and also with the University of Ottawa, ON K1N 6N5, Canada (E-mail: haiwei.dong@ieee.org).}%
\thanks{Marc St-Hilaire is with the School of Information Technology, Department of Systems and Computer Engineering, Carleton University, Ottawa, ON, K1S 5B6, Canada (E-mail: marc\_st\_hilaire@carleton.ca).}%
\thanks{Abdulmotaleb El Saddik is with the Multimedia Communications Research Laboratory (MCRLab), University of Ottawa, Ottawa, ON K1N 6N5, Canada (E-mail: elsaddik@uottawa.ca).}
\thanks{(Corresponding author: Haiwei Dong.)}}



\markboth{IEEE Transactions on Mobile Computing}%
{Shell \MakeLowercase{\textit{et al.}}: A Sample Article Using IEEEtran.cls for IEEE Journals}

\IEEEpubid{}

\maketitle

\begin{abstract}
The continuous evolution of cloud gaming has enabled users to play high-quality cloud games on thin clients, such as smartphones and tablets. However, how to effectively transmit the gameplay frames with high resolution and low latency, especially over wireless networks, remains a significant challenge. To address this challenge, we propose a Deep Reinforcement Learning (DRL)-based Interest-aware Adaptive Bitrate (ABR) scheme, called DIA, that guarantees the quality of perceptually key objects in each frame, while dynamically allocating lower resolutions for non-key objects to conserve bandwidth under constrained network conditions. By leveraging an off-the-shelf semantic segmentation model, the proposed DIA scheme can adjust the bitrate at the object-level. We also design a transformer-based feature extractor and an interest-aware reward function to enable the proposed DIA scheme to identify key objects and dynamically adjust their bitrates. This adaptation is guided by users' in-game movement speed, object features, and prevailing network conditions and operates without requiring additional hardware, such as gaze trackers. Extensive experiments show that the proposed DIA scheme outperforms existing ABR schemes in dynamic wireless network environments. Notably, DIA achieves at least a 20\% higher normalized average Quality of Experience (QoE) than the State-Of-The-Art (SOTA) ABR algorithms in 5G scenarios.
\end{abstract}


\section{Introduction}
\IEEEPARstart{T}{he} emergence of cloud gaming has enabled users to play high-quality games on thin clients, such as smartphones, tablets, and lightweight consoles, by offloading computationally intensive tasks, such as scene rendering and game logic computations, on cloud servers \cite{cloudgaming}. Thus, clients only need to decode the received gameplay frames, thereby significantly reducing hardware requirements on the user side. This approach eliminates the need for expensive GPUs and higher-performance CPUs, enabling cost-effective access to ``Triple-A (AAA)" games via affordable devices. Hence, leading technology companies have aggressively invested in this industry, launching commercial cloud gaming platforms such as Sony’s PlayStation Now \cite{ps}, NVIDIA GeForce Now \cite{nvidia}, and Microsoft Cloud Gaming \cite{xbox}. 

Moreover, recent advancements in wireless communication technologies, e.g., 5G, WiFi 7, have further accelerated the growth of cloud gaming by enabling ubiquitous and continuous Internet connectivity. However, maintaining high Quality of Experience (QoE) under fluctuating wireless environments remains a significant challenge, due to its strict latency, reliability, and bandwidth requirements. 



To overcome the above challenge, Adaptive Bitrate (ABR) schemes have been widely used in regular video streaming services. These schemes can dynamically adjust the encoding bitrate for each video chunk (typically lasting several seconds) based on the estimated network conditions, such as network throughput and buffer occupancy \cite{chunk}. Compared to conventional video streaming, cloud gaming imposes stricter requirements on bandwidth, latency, jitter, and packet loss. For example, playing Cyberpunk 2077 at 4K resolution and 120 Frames Per Second (FPS) demands a minimum bandwidth of 45 Mbps \cite{nvidia_systemreqs}, while watching 4K videos on Netflix requires only 15 Mbps \cite{netflix}. The latency requirement of cloud gaming is typically less than 50 ms \cite{nsdi24}, while regular video transmissions can tolerate second-level latencies due to the buffering mechanisms. Furthermore, unlike regular video streaming, which can be pre-encoded into multiple resolution versions for selection by ABR schemes \cite{adaptive}, cloud gaming cannot rely on pre-encoding due to its need for real-time rendering of interactive content \cite{deepgame}. Hence, the ABR schemes must dynamically adjust the bitrate for the gameplay scenes based on real-time network conditions and user interactions. However, most existing ABR schemes are at the chunk level, where adjustments are made over chunks that can last several seconds, as each chunk contains multiple frames. This coarse-grained approach is not suitable for cloud gaming, where low latency is critical for maintaining optimal QoE \cite{Nebula}. Fine-grained bitrate allocation is needed to better adapt to the dynamic nature of gameplay and user interactions. Additionally, existing ABR schemes ignore the fact that players typically focus on a limited set of objects within each frame, such as vehicles, obstacles, and enemies, rather than the entire scene. Therefore, allocating an equal bitrate to all objects within a frame leads to unnecessary bandwidth usage, which in turn degrades the QoE for cloud gaming users.






To address this limitation, we propose a \textbf{D}eep Reinforcement Learning (DRL)-based \textbf{I}nterest-aware \textbf{A}BR (DIA) scheme that dynamically prioritizes the perceptual quality of \textbf{key objects} (defined as those that attract user attention during gameplay) under constrained bandwidth conditions. By leveraging an off-the-shelf semantic segmentation model, the proposed DIA scheme performs object-level bitrate allocation, enabling more efficient and targeted transmission decisions. Since cloud gaming is a highly interactive application, object motion speed and user behavior, particularly in-game movement speed, play a crucial role in identifying key objects. Therefore, by jointly considering the user's in-game movement speed, object features, and network conditions, the proposed DIA scheme can identify key objects and guarantee their visual quality, while allocating lower resolutions for \textbf{non-key objects} (i.e., the remaining objects within the same frame) to reduce overall bandwidth consumption. As a result, the users' QoE can be maintained even under dynamic network conditions. Different from general content-aware ABR schemes that mainly adjust the bitrate at the chunk or frame level according to content complexity, DIA is specifically designed for latency-sensitive cloud gaming by exploiting interaction-driven object-level attention, where users typically focus on a limited set of key objects within each frame.


The main contributions of our work are summarized as follows:
\begin{itemize}
    \item We formulate a cloud-gaming-specific object-level ABR problem, where resolution decisions are made for individual in-game objects rather than entire chunks or frames. Based on this formulation, we design DIA to preserve the quality of key objects while reducing the bitrate of non-key objects under constrained network bandwidth.
    
   
    \item We design an interest-aware DRL framework for object-level bitrate allocation under thin-client and low-latency cloud gaming constraints. The state, action, and reward are designed to fit cloud gaming by jointly considering object semantic labels, object-level features, user in-game movement speed, latent inter-object relationships, and network conditions.


    \item We introduce new QoE metrics to better evaluate user experience in cloud gaming by considering latency and the quality of perceptually key objects. This is particularly important given cloud gaming’s stringent latency requirements and its highly interactive nature, where users' attention is often focused on a limited set of key objects.
   
    \item Extensive experiments demonstrate that the proposed approach significantly improves users' QoE compared to State-Of-The-Art (SOTA) ABR schemes, especially in dynamic wireless network environments.
\end{itemize}

The remainder of this paper is organized as follows: The related work is introduced in Section \uppercase\expandafter{\romannumeral2}. We present the system model in Section \uppercase\expandafter{\romannumeral3}. In Section \uppercase\expandafter{\romannumeral4}, we elaborate details of the proposed user interest-aware DRL-based ABR scheme. The simulation settings and results are shown in Section \uppercase\expandafter{\romannumeral5}. In Section \uppercase\expandafter{\romannumeral6}, we conclude the paper and outline future research directions.

\section{Related Work}
In this section, we introduce the recent research on ABR schemes, including rule-based ABR schemes and learning-based ABR schemes. We also briefly present the efforts on identifying key objects and Region-of-Interest (ROI) in frames, which can be used to improve the performance of ABR schemes.

\subsection{Rule-based ABR Schemes}
Existing ABR schemes \cite{ABRsurvey} can be categorized as rule-based and learning-based schemes. Rule-based ABR schemes are designed based on fixed control rules for making bitrate decisions by considering estimated network throughput, buffer occupancy, and rebuffering events. For instance, FESTIVE \cite{FESTIVE} makes bitrate selections based on the estimated bandwidth. On the other hand, Huang \emph{et al}. \cite{BBA} proposed a Buffer-Based ABR (BBA) scheme that selects bitrate based on the buffer threshold without bandwidth estimation or historical data analysis. Another buffer-based ABR scheme (named BOLA) that relies on buffer occupancy to adjust the bitrate is proposed in \cite{bola}. The difference between BBA and BOLA is that BOLA leverages Lyapunov optimization to maximize the selected bitrate while minimizing the rebuffering time, rather than relying on thresholds. Model Predictive Control (MPC) is another widely adopted class of scheme for bitrate selections under varying network conditions. The SOTA MPC-based scheme is RobustMPC \cite{RobustMPC}, which attempts to predict the throughput, and combines the predicted throughput and buffer occupancy to solve a QoE optimization problem for bitrate selections.

\subsection{Learning-based ABR Schemes}
Artificial Intelligence (AI) models are emerging recently and have already been utilized to improve the performance of ABR schemes. Since the bitrate selection can be modeled as a Markov Decision Process (MDP), most of the existing learning-based ABR schemes adopt DRL models to make bitrate selection decisions. For example, Pensieve \cite{Pensieve} is an SOTA DRL-based ABR scheme. It makes the bitrate selection decisions for the next chunk based on the current observations (including throughput, buffer occupancy, and video chunk sizes). Authors in \cite{stick} proposed a new DRL-based ABR scheme, named Stick, which fuses the DRL-based and buffer-based methods. Compared to Pensieve, Stick does not directly generate bitrate selections; it outputs buffer-bound to control the buffer-based method to adjust bitrate. Wu \emph{et al}. \cite{SR-ABR} proposed SR-ABR, which integrates a Super-Resolution (SR) technique on the client side to enhance video quality. In this way, the users' QoE can still be maintained even when only low bitrate videos are received. ZGming \cite{zgaming} is another client-side scheme. It can significantly reduce the latency by predicting the game frames on the client side when the network conditions are poor.



Although conventional DRL-based ABR schemes outperform rule-based ABR schemes, some researchers argue that the conventional DRL-based schemes may fail to obtain optimal strategies in unseen or fluctuating conditions. To overcome this issue, several advanced DRL-based approaches have been proposed. For instance, a self-play Reinforcement Learning (RL) approach (named as Zwei) is proposed in \cite{zwei}. Unlike conventional DRL-based ABR schemes, Zwei trains its policies through self-competition of trajectories, which makes Zwei to more effectively meet the actual requirement. Similarly, Li \emph{et al}. \cite{audio} proposed a self-play RL-based ABR scheme that jointly selects bitrates for both video and audio content. Unlike existing self-play RL-based ABR schemes, the authors in \cite{optimizing} consider the scene complexity and model the bitrate selection problem as an explicit objective. Meta-learning \cite{MetaABR, A2BR, Ahaggar} is another promising model to address the deployment challenges of conventional DRL-based ABR schemes, as it leverages experience and meta-data from previous tasks to quickly adapt to new tasks. MetaABR uses meta-learning to improve the performance of the DRL model by jointly training multiple learning tasks with a shared meta-critic, which can provide transferable knowledge to assist bitrate selection across tasks. Similarly, Ahaggar \cite{Ahaggar} also leverages meta-learning to improve the performance of RL in heterogeneous environments. The key difference is that Ahaggar implements the learning scheme on the server side, which provides quality-aware bitrate guidance to clients running their own heuristic algorithms.

In addition to improving adaptability, recent ABR studies have also considered the perceptual quality to design ARB schemes. Tang \emph{et al}. \cite{360} found that the relationship between video quality and bitrate is non-linear, and proposed a new QoE metric based on structural similarity. By using the new QoE metric to design the reward function of the proposed DRL-based scheme, they can achieve a higher QoE performance compared to existing schemes. The authors in \cite{Successor} first designed a QoE preference analysis scheme to infer the personal QoE preference of a user. Based on these preferences, their DRL-based scheme can adjust the reward setting to maximize the user's QoE.


\subsection{Identifying Key Objects}
The direct way to identify key objects in each frame is to use gaze trackers \cite{deepfovea}. Gaze trackers can provide detailed information about where users are looking at in each frame, which can be treated as the key objects. However, not all clients are equipped with gaze trackers. Therefore, indirect approaches are proposed to address this issue. For example, several efforts have been made to identify key objects by considering the objects' colour, contrast, luminance, texture, etc., in frames \cite{15, 107}.

Moreover, more recent works have leveraged AI models to identify key objects in frames. For instance, the DRL model is used in \cite{intelligent-DRL} to determine the size and quality of ROI and non-ROI in each frame based on the expected throughput. Cai \emph{et al}. \cite{novel} leveraged YOLOv3 to design an object detector to identify ROI, but their approach only identifies objects, not their relative importance. Mossad \emph{et al}. \cite{deepgame} also utilized YOLOv3 for detecting objects, followed by an Long-Short Term Network (LSTM) to predict the ROI. Similarly, Wang \emph{et al}. \cite{map} adopted a semantic segmentation model to identify the ROI. These approaches, though effective in object detection, still rely on pre-defined rules to determine which objects or areas are user-interested. T-Gaming \cite{T-gaming} introduced a Human Visual System (HVS) to distinguish user-interested regions based on visual saliency. Yet, real gameplay scenarios are influenced by multiple factors beyond saliency, such as in-game movement speed, semantic labels, and network conditions.


\subsection{Summary}
The rule-based ABR schemes are lightweight and simple to deploy. However, their performance is highly impacted by the settings of parameters, which require manual tuning. In contrast, the learning-based ABR scheme can overcome this issue by automatically learning the optimal bitrate decisions in given network environments. The recent efforts in learning-based ABR schemes can be categorized into two directions: 1) improving the generalization ability of ABR schemes; 2) considering a more accurate QoE metric to improve the efficiency of ABR schemes. However, most of the existing ABR schemes are at the chunk level, which consists of multiple frames. This coarse-grained method ignores the fact that human attention is often focused on certain key objects (e.g., vehicles, people) in a frame, especially in interactive scenarios like cloud gaming. Therefore, the existing ABR schemes may waste bandwidth on perceptually unimportant regions and fail to maximize QoE under constrained network conditions.

Regarding to key object identification, existing approaches either require additional hardware (e.g., eye-tracking sensors) to track user attention, or rely on pre-defined rules to determine which objects or regions are perceptually important within a frame. Hence, how to automatically identify key objects in frames and leverage them to improve the performance of ABR schemes remains a significant challenge.

\section{System Model}
In this section, we demonstrate the mathematical modeling for the proposed DIA scheme. Table \ref{tab:notation} summarizes the notations used in this section.

\begin{table}[!ht]
\centering
\caption{Notation Summary for System Model}
\label{tab:notation}
\begin{tabular}{@{}lp{6cm}@{}}
\toprule
\textbf{Symbol} & \textbf{Description} \\
\midrule
\multicolumn{2}{c}{\textit{A. Network Model}} \\
\midrule
$\mathcal{U}$ & Set of users: $\{u_1, u_2, \ldots, u_N\}$ \\
$u_n$ & $n$-th user in the system \\
$bw_t(u_n)$ & Available bandwidth for user $u_n$ at time slot $t$ \\
$BW$ & Total channel bandwidth \\
$P_t$ & Transmit power at time $t$ \\
$G_t, G_r$ & Transmit and receive antenna gains \\
$N_0$ & Noise power \\
$d(t)$ & Distance between BS and user at time $t$ \\
$\Lambda$ & Path loss exponent \\
$\psi(v_t(u_n))$ & Doppler degradation factor for user speed $v_t(u_n)$ \\
$v_t(u_n)$ & User $u_n$'s speed at time $t$ \\
\midrule
\multicolumn{2}{c}{\textit{B. User Mobility Model}} \\
\midrule
$(x_t(u_n), y_t(u_n))$ & User $u_n$'s position coordinates at time $t$ \\
$v_{\max}$ & Maximum user speed \\
$\theta_t(u_n)$ & User movement direction angle at time $t$ \\
$\Delta t$ & Time interval between position updates \\
$(x_{BS}, y_{BS})$ & Base station coordinates \\
\midrule
\multicolumn{2}{c}{\textit{C. Content Model}} \\
\midrule
$\mathcal{F}_t$ & Set of frames in time slot $t$: $\{f_{t,1}, \ldots, f_{t,M}\}$ \\
$T$ & Total gameplay duration \\
$M$ & Number of frames per time slot \\
$\mathcal{O}$ & Set of all game objects: $\{o_1, \ldots, o_K\}$ \\
$K$ & Total number of objects \\
$\mathcal{O}(f_{t,m})$ & Objects in frame $f_{t,m}$ ($\subseteq \mathcal{O}$) \\
$o_k$ & Object tuple: $(l_k, s_t(o_k), vg_t(o_k))$ \\
$l_k$ & Semantic label of object $o_k$ \\
$s_t(o_k)$ & Pixel area of object $o_k$ at time $t$ \\
$vg_t(o_k)$ & In-game motion speed of object $o_k$ at time $t$ \\
$(x_t(o_k), y_t(o_k))$ & Position coordinates of object $o_k$ \\
$r(o_k)$ & Resolution level assigned to object $o_k$ \\
$br_t(o_k)$ & Bitrate of object $o_k$ at time $t$ \\
$br_t(f_{t,m})$ & Total bitrate of frame $f_{t,m}$ at time $t$ \\
\midrule
\multicolumn{2}{c}{\textit{D. User Interest Model}} \\
\midrule
$p_k(t)$ & Interest score of object $o_k$ at time $t$ \\
$I(\cdot)$ & User interest function \\
$vg_t(u_n)$ & User $u_n$'s in-game movement speed at time $t$ \\
\midrule
\multicolumn{2}{c}{\textit{E. Transmission Model}} \\
\midrule
$L(\mathcal{F}_t)$ & Total end-to-end latency for frames $\mathcal{F}_t$ \\
$L(f_{t,m})$ & Latency for frame $f_{t,m}$ \\
$L_{\text{tran}}$ & Transmission latency \\
$L_{\text{prop}}$ & Propagation latency \\
$L_{\text{que}}$ & Queuing latency \\
$L_{\text{proc}}$ & Processing latency \\
$c$ & Signal propagation speed \\
$b_s$ & Buffer size \\
$L_{\text{buf}}$ & Buffer-induced latency \\
$f_r$ & Frame rate \\
$M_{\text{buf}}$ & Number of buffered frames \\
$L_{\text{reb}}(\mathcal{F}_t)$ & Rebuffering time for frames $\mathcal{F}_t$ \\
$br_{\text{play}}(\mathcal{F}_t)$ & Bitrate of played frames in slot $t$ \\
$br_{\text{recv}}(\mathcal{F}_t)$ & Bitrate of received frames in slot $t$ \\
\midrule
\multicolumn{2}{c}{\textit{F. QoE Model}} \\
\midrule
$QoE(\mathcal{F}_t)$ & QoE value for user $u_n$ at time $t$ \\
$Q_{br}(o_k,t)$ & Bitrate quality function of object $o_k$ \\
$BR_{var}(o_k,\Delta t)$ & Bitrate variation of object $o_k$ between slots \\
$\alpha, \beta, \gamma, \theta$ & Weights for QoE components \\
\bottomrule
\end{tabular}
\end{table}

\subsection{Network Model}

\begin{figure}[!t]
\centering
\includegraphics[width=\linewidth]{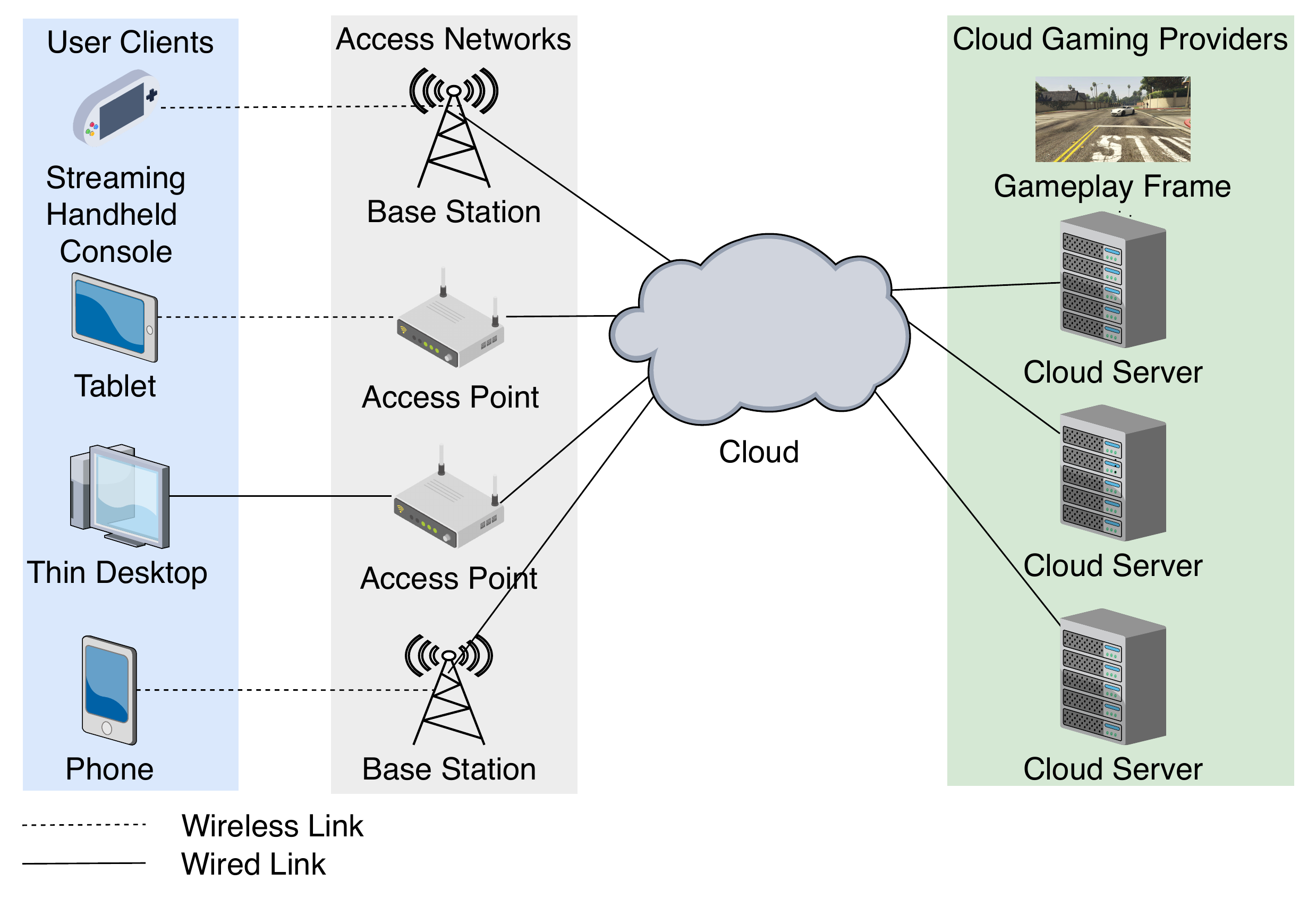}
\caption{Wireless cloud gaming network model.}
\label{network-model}
\end{figure}

In this paper, we consider a wireless cloud gaming scenario in which users receive gameplay frames from a cloud server. All frame rendering and encoding tasks are performed at the cloud server. After the frames are rendered and encoded, they are transmitted to user devices through Base Stations (BSs) or Access Points (APs) over time-varying wireless networks. Consider a set of $N$ users, $\mathcal{U} = \{u_1, u_2, \dots, u_n, \dots, u_N\}$, the bandwidth of user $u_n$ at the beginning of time slot $t$ can be denoted as $bw_t(u_n)$. The assumed scenario is illustrated in Fig. \ref{network-model}.

\subsection{User Mobility Model}
The user mobility model is used for calculating the $d(t)$ and $v_t(u_n)$ at each time slot. In this paper, we assume that each user $u_n$ is randomly moving in a 2D space. The user's position at time $t$ is denoted as $(x_t(u_n), y_t(u_n))$, where $x_t(u_n)$ and $y_t(u_n)$ are the horizontal and vertical coordinates of user $u_n$ at time $t$, respectively. To represent diverse user movement scenarios, we model the user's speed $v_t(u_n)$ as a random variable sampled from a uniform distribution in the range $[0, v_{\max}]$, where $v_{\max}$ denotes the maximum speed of the user. This allows the model to capture a wide range of mobility patterns, from stationary users to those moving at high speed. 

\subsection{Content Model}
In cloud gaming, gameplay content is transmitted as a sequence of frames, where each frame includes multiple objects. We assume that the gameplay session frames are transmitted in discrete time slots $t$, indexed by $t = 1, 2, \dots, T$, where $T$ is the total duration of the gameplay session in seconds. The set of frames transmitted within time slot $t$ is denoted by $\mathcal{F}_t = \{f_{t,1}, f_{t,2}, \dots,f_{t, m} ,\dots, f_{t,M}\}$, where $M$ is the total number of frames within time slot $t$. Notably, if the duration of each time slot $t$ is 1 second, then $M$ represents the frame rate (FPS). We define $\mathcal{O} = \{o_1, o_2, \dots, o_k, \dots, o_K\}$ as the set of all possible objects in the gameplay session, where $K$ is the total number of objects. Note that any frame $f_{t,m}$ only contains a subset of $\mathcal{O}$, i.e., $\mathcal{O}(f_{t,m}) \subseteq \mathcal{O}$. We define objects that users are focused on as \textbf{key objects}, other objects within the same frame are defined as \textbf{non-key objects}. Each object $o_k \in \mathcal{O}(f_{t,m})$ can be further represented as a tuple $o_k = (l_k, s_t(o_k), vg_t(o_k))$, where $l_k$ is the semantic label (e.g., car, pedestrian), $s_t(o_k)$ stands for its pixel area in the frame, and $vg_t(o_k)$ represents its in-game motion speed at time $t$. Moreover, $vg_t(o_k)$ can be calculated as the Euclidean distance between the current position and the previous position of the object $o_k$ at time $t$:  
\begin{equation}
vg_t(o_k) = \sqrt{ \left( x_t(o_k) - x_{t-1}(o_k) \right)^2 + \left( y_t(o_k) - y_{t-1}(o_k) \right)^2 }
\end{equation}
where $x_t(o_k)$ and $y_t(o_k)$ are the horizontal and vertical coordinates of the centroid position of object $o_k$ at time $t$, respectively.

Notably, when it comes to cloud gaming, users usually pay attention to only a few key objects per frame, such as vehicles, obstacles, and enemies. Therefore, assigning a uniform bitrate to all objects in a frame may lead to unnecessary bandwidth consumption. To address this issue, we propose an object-level ABR scheme, which dynamically selects different bitrates for different objects in a frame based on the user's interest. For each object $o_k$ in frame $f_{t,m}$, we define a discrete resolution level $r(o_k) \in \mathcal{R}_{\text{es}}$, where $\mathcal{R}_{\text{es}} = \{r_1, r_2, \dots, r_l, \dots, r_L\}$ is the set of available resolution levels (e.g., 480p, 720p, 1080p, 2k, 4k). After encoding, each resolution level $r(o_k)$ is mapped to a specific encoded bitrate $br_t(o_k)$. Hence, the total bitrate of $f_{t,m}$ can be calculated as:
\begin{equation}
br_t(f_{t,m}) = \sum_{o_k \in O(f_{t,m})} br_t(o_k)
\end{equation}
where $br_t(f_{t,m})$ is the total bitrate of frame $f_{t,m}$. Consequently, determining the objects that users are most interested in becomes a critical step in our object-level ABR scheme.

\subsection{User Interest Model}
In general, users' interest in objects within a frame can be influenced by various factors, such as the object's semantic label, pixel area, and movement speed \cite{gaze, saliency, map}. Notably, the importance of an object is not only determined by its individual features, but also by the gameplay scenario, which is formed by a combination of objects. For instance, a pedestrian may be a key object in a city scene with dense interactions, but it may become negligible in a highway scenario. This indicates that the latent relationships among objects are also important for identifying the key ones.

Moreover, the user's in-game movement speed also has a significant impact on their interest in objects. For example, if a user is driving a car at high speed in a racing game, s/he is more likely to focus on the road and nearby vehicles. However, if the user is driving at a lower speed, then s/he may also be interested in the surrounding environment, such as pedestrians and buildings.


Therefore, we denote $p_k(t)$ as the interest score of object $o_k$ at time $t$, which can be calculated by a user interest function $I(\cdot)$ as follows:
\begin{equation}
p_k(t) = I(l_k, s_t(o_k), p_t(o_k), vg_t(o_k), vg_t(u_n))   
\end{equation}
where $l_k$ is the semantic label of object $o_k$, $s_t(o_k)$ is the pixel area of object $o_k$ at time $t$, $p_t(o_k) = (x_t(o_k), y_t(o_k))$ represents the centroid position of object $o_k$ in the frame, $vg_t(o_k)$ is the in-game motion speed of object $o_k$ at time $t$, and $vg_t(u_n)$ is the user's in-game movement speed at time $t$. Notably, the impacts of the above factors on user interest are non-linear and complicated, which makes it extremely difficult to model the user interest function $I(\cdot)$ mathematically. Therefore, we leverage a neural network within the proposed DIA agent to estimate $p_k(t)$. Based on the estimated interest score $p_k(t)$, we can select the top-K interested objects in frame $f_{t,m}$ as the key objects.



\subsection{Transmission Model}
The total end-to-end latency $L(\mathcal{F}_t)$ of transmitting frames $\mathcal{F}_t$ from the cloud server to user $u_n$ at time $t$ can be modeled as:
\begin{flalign}
  L(\mathcal{F}_t) &= \sum_{m=1}^{M} L(f_{t, m}) \notag \\
  &= \sum_{m=1}^{M} \big[ L_{tran}(f_{t, m}) + L_{prop}(f_{t, m}) \notag \\
  &\qquad\quad + L_{que}(f_{t, m}) + L_{proc}(f_{t, m}) \big] \label{latency_e2e} \\
  L_{trans}(f_{t, m}) &= \frac{br_t(f_{t, m})}{bw_t(u_n)} \\
  L_{prop}(f_{t, m}) &= \frac{D(t)}{c}
\end{flalign}
where $L_{trans}(f_{t, m})$ is the transmission latency of frame $f_{t, m}$, $L_{prop}(f_{t, m})$ is frame $f_{t, m}$'s propagation latency, $D(t)$ is the distance between the cloud server and the user device at time $t$, $c$ is the signal propagation speed in physical links, $L_{que}(f_{t, m})$ and $L_{proc}(f_{t, m})$ are the queuing latency and processing latency of frame $f_{t, m}$, respectively.
To satisfy the low latency requirement of cloud gaming, the relationship between the bandwidth and the bitrate of frames $\mathcal{F}_t$ at time $t$ can be modeled as:
\begin{equation}
bw_t(u_n) \geq \frac{1}{L_{max}} \cdot \sum_{m=1}^{M} br_t(f_{t,m}) \label{bw}
\end{equation}
where $L_{max}$ is the maximum end-to-end latency that does not degrade users' QoE. 

Since user $u_n$'s available bandwidth $bw_t(u_n)$ may vary over time due to the dynamic nature of wireless networks, which means that $bw_t(u_n)$ may not always satisfy Eq. \eqref{bw}. Therefore, to mitigate the impacts of the bandwidth fluctuations, a small buffer is utilized to store the received frames before playback. However, compared to regular video streaming, cloud gaming has more stringent latency requirements, which means the buffer size $b_s$ is much smaller than that of regular video streaming \cite{ACMMM}. As a result, rebuffering may occur more frequently in cloud gaming than in regular video streaming. The latency $L_{buf}(\mathcal{F}_t)$ that is caused by the implementation of the buffer can be modeled as:
\begin{align}
 L_{{buf}}(\mathcal{F}_t) = \frac{1}{f_r} \cdot M_{{buf}} \label{eq:buffer_latency} \\
\text{s.t.} \quad  \sum_{m=1}^{M_{{buf}}} br_t(f_{t,m}) \leq b_s \label{eq:buffer_constraint}
\end{align}
where $f_r$ is the frame rate, and $M_{buf}$ is the number of buffered frames. The constraint in Eq. \eqref{eq:buffer_constraint} ensures that the total bitrate of buffered frames does not exceed the buffer size $b_s$. Obviously, a larger buffer size $b_s$ means there are more buffered frames, which can increase $L_{buf}(\mathcal{F}_t)$.

To quantitatively evaluate the impact of rebuffering on users' QoE, we can calculate the rebuffering time $L_{reb}(\mathcal{F}_t)$ caused by transmitting frames $\mathcal{F}_t$ within time $t$ as follows:
\begin{align}
&L_{reb}(\mathcal{F}_t) = 
\max \left\{ 0, \frac{br_{play}(\mathcal{F}_t) - \bigl( br_{recv}(\mathcal{F}_t) + b_s \bigr)}{bw_t(u_n)} \right\} \\
 &br_{play}(\mathcal{F}_t) =  \sum_{m=1}^{M} br_t(f_{t,m}) \\
 &br_{recv}(\mathcal{F}_t) = \sum_{m_r=1}^{M_r} br_t(f_{t, m_r})
\end{align}
where $br_{play}(\mathcal{F}_t)$ is the total bitrate of frames that are played within time $t$, $br_{recv}(\mathcal{F}_t)$ is the total bitrate of frames that have been received by user $u_n$ within time $t$, $M_r$ is the total number of received frames, and $b_s$ is the buffer size. Additionally, the end-to-end latency $L(\mathcal{F}_t)$ can be rewritten as:
\begin{align}
  L(\mathcal{F}_t) &= L_{tran}(\mathcal{F}_t) + L_{prop}(\mathcal{F}_t) + L_{que}(\mathcal{F}_t) +\notag \\
         &\quad  L_{proc}(\mathcal{F}_t) + L_{buf}(\mathcal{F}_t) + L_{reb}(\mathcal{F}_t).
\end{align}

\subsection{QoE Model}

Unlike conventional ABR schemes that evaluate quality at the frame level, we model the QoE by aggregating the perceptual utility of individual key objects. Specifically, the QoE value for a frame sequence $\mathcal{F}_t$ is formulated as a weighted function of object-level bitrate quality, temporal smoothness, and system-level latency penalties:

\begin{align}
QoE(\mathcal{F}_t) &=  \sum_{m \in M} \Bigg[\sum_{k \in K} \left[ \alpha \cdot Q_{br}(o_k, t) - \beta \cdot L_{reb}(f_{t, m}) \right] -  \notag  \\
       &  \gamma \cdot BR_{var}(o_k, \Delta t)  - \theta \cdot L(f_{t, m}) \Bigg]\label{eq:QoE}
\end{align}
where $\alpha$, $\beta$, and $\gamma$ are the configurable weights, $Q_{br}(o_k, t)$ is the bitrate quality function of object $o_k$, $BR_{var}(o_k, \Delta t)$ is the bitrate variation of object $o_k$ between consecutive time slots, $L_{reb}(f_{t, m})$ is the rebuffering time of frame $f_{t, m}$, and $L(f_{t, m})$ is the end-to-end latency of frame $f_{t, m}$. 

This general QoE model is the foundation for designing the proposed new QoE metrics. By adopting different bitrate quality functions and weights, we can obtain different QoE metrics. More details can be found in Section \ref{sec:QoE}.


\section{DIA Scheme}
In this section, we present the details of the proposed DRL-based interest-aware ABR scheme. 

\subsection{Overview}
\begin{figure}[!t]
\centering
\includegraphics[width=3.0in]{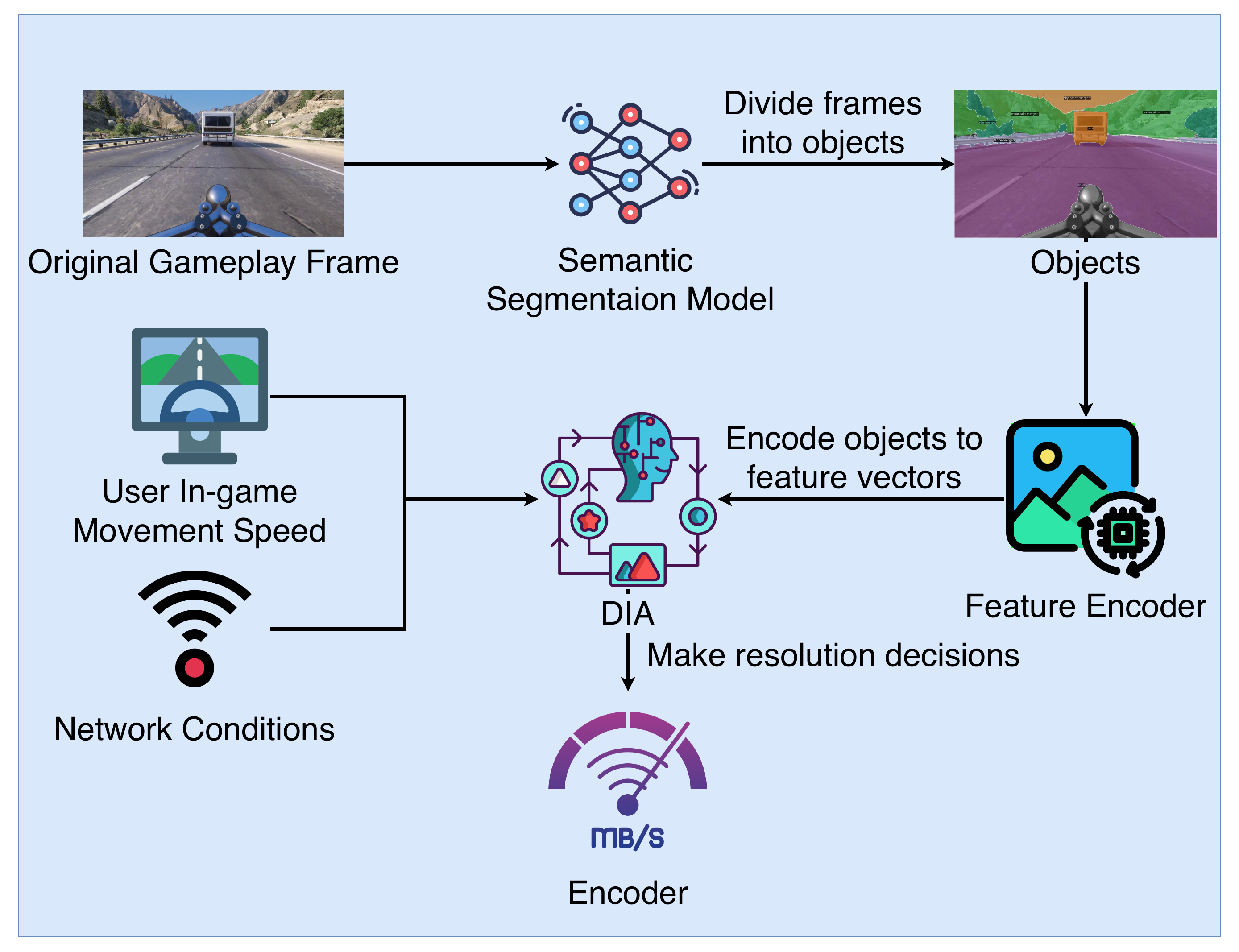}
\caption{Workflow of the DIA scheme.}
\label{workflow}
\end{figure}

To enable object-level bitrate allocation, we first apply an off-the-shelf semantic segmentation model (e.g., Mask2Former \cite{mask2former}) to segment each gameplay frame into multiple objects. Then, we utilize the Proximal Policy Optimization (PPO) model \cite{ppo} as the DRL model for the proposed DIA scheme. By taking the object features (including semantic labels, pixel areas, centroid positions, object motion information), user in-game movement speed, and network conditions as the inputs, the proposed DIA scheme can adjust the optimal resolution for each object in each frame. As a result, objects that contribute more to the overall QoE are implicitly assigned higher resolutions, whereas less important objects are allocated lower resolutions under the available bandwidth budget. The workflow of the proposed DIA scheme is illustrated in Fig. \ref{workflow}.

\subsection{Design of DIA Scheme}
Given the object features (including semantic label, object pixel area, and in-game motion speed), user in-game movement speed, and network conditions (such as buffer occupancy and available bandwidth) at each time slot $t$, the goal is to maximize the user's QoE by dynamically allocating the optimal bitrate for each object $o_k$ in each frame $f_{t,m}$ under dynamic network conditions. Semantic label and object pixel area can be obtained from the semantic segmentation model, while the in-game motion speed can be calculated based on the object's positional changes in consecutive frames. The user's in-game movement speed can be estimated by analyzing the change of background between consecutive frames. Lastly, network conditions can be obtained from the user's device. For simplicity, we don't assign a specific value to a user's in-game speed. Instead, we categorize it into three levels: low, medium, and high.

Therefore, the object-level bitrate selection problem can be formulated as a Markov Decision Process (MDP), which can be defined by the tuple $(\mathcal{S}, \mathcal{A}, \mathcal{P}, \mathcal{R})$, where $\mathcal{S}$ is the state space, $\mathcal{A}$ is the action space, $\mathcal{P}$ is the state transition probability, $\mathcal{R}$ is the reward set. 

To solve the formulated problem, we adopt a DRL model to learn the optimal policy $\pi^*(s)$ that maximizes the expected cumulative reward $\mathcal{R}$ over time. Since the state transition probability $\mathcal{P}$ is unknown in our scenario, only a model-free DRL model can be used to learn the optimal policy $\pi^*(s)$. We choose PPO as the DRL model due to its training stability and sample efficiency compared to other model-free models such as Deep Q-Network (DQN) and Actor-Critic (AC).

\subsubsection{State Space} 
The state $s_t \in \mathcal{S}$ is the observation that the DRL agent can obtain at time $t$. It includes the object-level features, user in-game movement speed, and network conditions. It can be represented as a vector:
\begin{equation}
s_t(o_k) = \left[ \mathbf{l}_k, \; \mathbf{s}_t(o_k), \; \mathbf{v}_t(o_k), \; \mathbf{n}_t^{\text{rem}}, \; \mathbf{bw}_t, \; \mathbf{b}_s^t \right]
\end{equation}
where $\mathbf{l}_k$ is the one-hot encoded semantic label of object $o_k$, $\mathbf{s}_t(o_k)$ is the normalized pixel area of object $o_k$ in frame $f_{t,m}$, $\mathbf{v}_t(o_k)$ is the normalized in-game motion speed of object $o_k$ at time $t$, $\mathbf{n}_t^{\text{rem}}$ is the percentage of remaining objects in frame $f_{t,m}$ that have not been assigned a bitrate yet, $\mathbf{bw}_t(u_n)$ is user $u_n$'s normalized available bandwidth at time $t$, and $\mathbf{b}_s^t$ is the normalized current buffer occupancy. The above one-hot encoding and normalizations are used to train the DRL model more stably and efficiently. 
\subsubsection{Action Space}
The action $a_t \in \mathcal{A}$ is the resolution selection made by the agent at time $t$, and $|\mathcal{A}|$ is the total number of available resolution levels. Specifically, the DRL agent selects a resolution level for each object $o_k$ in frame $f_{t,m}$. Thus, the action space can be defined as a discrete set of resolution levels $\mathcal{A} =\mathcal{R}_{\text{es}}= \{320 \times 240,\ 854 \times 480,\ 1280 \times 720,\ 1920 \times 1080,\ 2560 \times 1440,\ 3840 \times 2160\}$. The action can be represented as:
\begin{equation}
a_t(o_k) = r_l \in \mathcal{R}_{\text{es}}.
\end{equation}
Notably, the resolution level only represents the relative encoding quality for individual objects rather than the entire frame. Once the action is selected, the bitrate of object $o_k$ can be calculated based on the selected resolution level $r_l$.

\subsubsection{Interest-aware Reward Function}
To maximize the user's QoE, the agent needs to identify the key objects and assign higher bitrates to them, especially when the available bandwidth is limited. To achieve this goal, we design an interest-aware reward function that evaluates the agent's actions based on whether it successfully allocates higher resolutions for key objects when the available bandwidth is insufficient. 

We leverage a human-labeled dataset~\cite{dataset}, in which the key objects in each gameplay clip are annotated by human annotators, providing ground-truth interest labels at the object level. Specifically, if the action $a_t(o_k)$ allocates a higher bitrate for a key object $o_k$, the agent receives a positive bitrate reward; otherwise, it receives a penalty. This design encourages the agent to prioritize key objects and allocate higher resolutions to them when available bandwidth is constrained. The overall reward function is defined as follows:


\begin{align}
  \mathcal{R}(s_t, a_t) =& \omega_1 \cdot \mathcal{R}_{br}(g_k, a_t) + \omega_2 \cdot \mathcal{R}_{reb}(t) + \\ \notag
  & \omega_3 \cdot \mathcal{R}_{var}(t) + \omega_4 \cdot \mathcal{R}_{lat}(t) \label{reward_function}
\end{align}
where $\mathcal{R}_{br}(g_k, a_t)$ is the bitrate reward ($g_k \in \{0, 1\}$ denotes the ground-truth interest label of object $o_k$). $\mathcal{R}_{reb}$, $\mathcal{R}_{var}$, and $\mathcal{R}_{lat}$ represent the penalties for rebuffering, bitrate variation, and latency, respectively. 

To ensure reproducibility and training stability, the specific calculations are defined as follows:
\paragraph{Bitrate Reward ($\mathcal{R}_{br}$)}
Since resolution levels are discrete, we use a look-up table to map the resolution and interest label to a scalar score, as detailed in Table~\ref{tab:resolution_table}.

\paragraph{Normalized Penalties}
To improve the numerical stability of DRL training, the penalty terms are normalized and clipped to the range $[0,1]$ using a function 
$\mathcal{C}(x)=\min(\max(x,0),1)$. 
Different from the bitrate reward, which is determined by the selected object-level resolution and the ground-truth interest label, the rebuffering, latency, and bitrate-variation penalties are obtained from the feedback of the network transmission environment after executing the selected action. 
Specifically, under the current bandwidth $bw_{t,m}(u_n)$, buffer size $b_s$, and selected bitrate allocation, the environment returns the cumulative rebuffering time $L_{reb}(\mathcal{F}_t)$ of frames $\mathcal{F}_t$, the total end-to-end latency $L(\mathcal{F}_t)$, and the aggregated bitrate variation $BR_{var}(\mathcal{F}_t)$. These feedback values correspond to the rebuffering, latency, and bitrate variation formulated in Section~III, while the equations in Section~III define their system-level meanings and dependent variables.

For instance, the rebuffering penalty is calculated as
\begin{equation}
\mathcal{R}_{reb}(t)=
\mathcal{C}\left(
\frac{L_{reb}(\mathcal{F}_t)}{\tau_{reb}}
\right),
\end{equation}
where $\tau_{reb}$ is a scaling factor. 
Similarly, the latency and bitrate-variation penalties are calculated as
\begin{align}
\mathcal{R}_{lat}(t) &=
\mathcal{C}\left(
\frac{L(\mathcal{F}_t)}{\tau_{lat}}
\right), \\
\mathcal{R}_{var}(t) &=
\mathcal{C}\left(
\frac{BR_{var}(\mathcal{F}_t)}{\tau_{var}}
\right),
\end{align}
where $\tau_{lat}$ and $\tau_{var}$ are the corresponding scaling factors. 
The scaling factors are determined according to the practical ranges or upper bounds of the corresponding feedback values in the simulation environment.

In our experiments, the weighting parameters are set to $\omega_1=2.0$, $\omega_2=-4.0$, $\omega_3=-0.2$, and $\omega_4=-1.0$. These values were determined through empirical tuning to balance the magnitude of each component. The scaling factors are empirically set to $\tau_{reb}=10$, $\tau_{lat}=5$, and $\tau_{var}=5$ to ensure numerical stability during training.

\begin{table}[htbp]
\caption{Discrete Resolution Rewards ($\mathcal{R}_{br}(g_k, a_t)$)}
\label{tab:resolution_table}
\centering
\begin{tabular}{c|cc}
\hline
\textbf{Resolution ($a_t$)} & \textbf{Key Object ($g_k=1$)} & \textbf{Non-Key Object ($g_k=0$)} \\ \hline
3840 $\times$ 2160  & 10.0 & 3.0 \\ 
2560 $\times$ 1440 & 5.0 & 1.5 \\ 
1920 $\times$ 1080  & 3.0 & 0.5 \\ 
1280 $\times$ 720  & 0.5 & 0.0 \\ 
854 $\times$ 480  & -2.0 & -1.0 \\ 
320 $\times$ 240  & -5.0 & -3.0 \\ \hline
\end{tabular}
\end{table}

\subsection{Neural Network Architecture}
\begin{figure}[!t]
\centering
\includegraphics[width=2.0in]{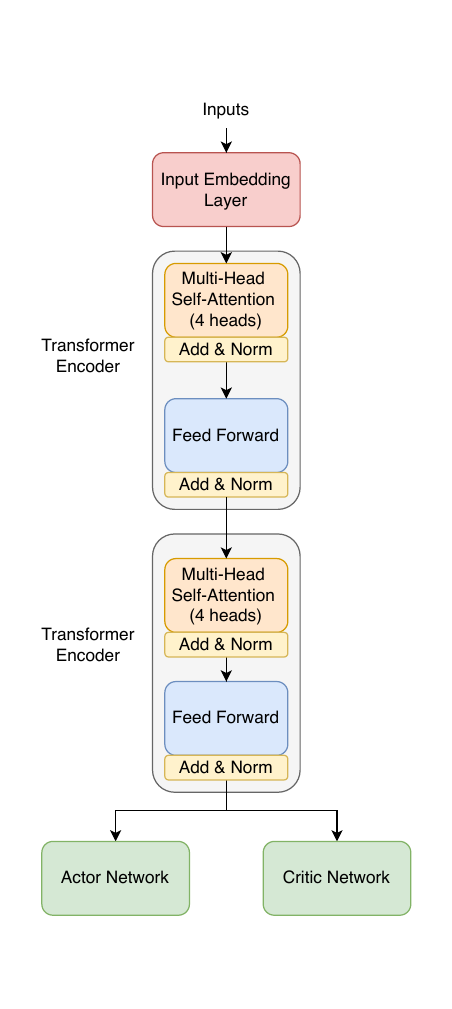}
\caption{The neural network architecture of the DRL model, which includes a transformer-based feature extractor and an actor-critic architecture.}
\label{drn_architecture}
\end{figure}

To meet the low latency requirements of cloud gaming, we design a lightweight DRL model that includes two lightweight neural network architectures: a transformer-based feature extractor and an actor-critic architecture. The transformer-based feature extractor is used to capture the latent relationships among objects within the same frame, and its outputs are fed to the actor-critic architecture to learn the optimal policy for bitrate allocation based on the learned latent relationships, user in-game movement speed, and network conditions. The overall architecture of the DRL model is illustrated in Fig. \ref{drn_architecture}.

\subsubsection{Transformer-based Feature Extractor}

It consists of a linear embedding layer followed by two stacked transformer encoder blocks, where each encoder block includes a multi-head self-attention mechanism with four heads, a feed-forward network, a normalization layer, and residual connections. The Transformer-based feature extractor is shared by both the actor and critic networks and is jointly trained via backpropagation from both the policy and value function losses.


\subsubsection{Actor-Critic Architecture}

The actor and critic networks share a similar architecture. Both consist of three fully connected layers, each layer followed by a ReLU activation. A normalized layer is applied after the second ReLU to stabilize training. The actor network then uses an additional fully connected layer followed by a Softmax activation to generate a probability distribution over discrete resolution levels, while the critic network utilizes a linear layer to estimate the scalar state-value.

\subsection{Training and Practical Implementation}

In this subsection, we present the details for training and practical implementation of the proposed DIA scheme.

\subsubsection{Dataset}
We collected gameplay clips from GTA V, a popular open-world game that includes a wide variety of scenes and object types, such as urban and rural environments, roads, vehicles, and pedestrians. It also contains diverse gameplay patterns, including driving, shooting, and action, etc. This diversity improves the generalization of the trained model, allowing it to adapt to various cloud gaming scenarios. The dataset includes around 500 gameplay clips and more than 1500 gameplay images. Five independent human annotators annotate the key objects. To ensure reliability and handle subjective variability, we adopted a multi-interest aggregation strategy, where objects consistently identified by multiple annotators are designated as ground-truth key objects. This provides robust interest labels at the object level. There are 22 pre-defined object classes active in the collected dataset \cite{dataset}. Notably, these class definitions are generated from the standard COCO dataset \cite{coco} (e.g., mapping specific game objects to general classes like 'Person', 'Vehicle', 'Building'). This ensures that the state space definition is generic and transferable to other realistic game genres.


\subsubsection{Hardware Environment}
The training is conducted on a Linux server equipped with dual Intel Xeon Platinum 8352V CPUs and 4 RTX 4090 GPUs (each with 24 GB), using Python 3.12.2 and PyTorch 2.2.2 on Ubuntu 22.04.

\subsubsection{Hyperparameter Settings}
We summarize the key hyperparameters used in the training of our PPO-based DRL model in Table~\ref{tab:hyperparameters}. Additionally, to encourage the agent to explore more at the beginning of the training and gradually shift to exploitation, we apply Eq. \eqref{eq:entropy_weight_decay} to exponentially decay the entropy weight $\delta$ over the training episodes. 
\begin{equation}
\eta_{epi} = \max\left(\eta_{final},\ \eta_{init} \cdot \delta ^{epi}\right)
\label{eq:entropy_weight_decay}
\end{equation}
where $epi$ is the current training episode.

\begin{table}[!t]
\centering
\caption{Summary of Hyperparameter Settings}
\label{tab:hyperparameters}
\begin{tabular}{|l|c|}
\hline
\textbf{Parameter} & \textbf{Value} \\
\hline
Total training episodes ($H$) & 5000 \\
Batch size ($B$) & 256 \\
Learning rate ($l_r$) & $1 \times 10^{-6}$ \\
Discount factor ($\phi$) & 0.99 \\
PPO epochs ($P_e$) & 8 \\
Clipping parameter ($\epsilon$) & 0.2 \\
GAE parameter ($\lambda$) & 0.95 \\
Max gradient norm ($\| \nabla \|_{\text{max}}$) & 0.4 \\
Initial entropy weight ($\eta_{\text{init}}$) & 0.1 \\
Final entropy weight ($\eta_{\text{final}}$) & 0.01 \\
Entropy decay rate ($\delta$) & 0.999 \\
Number of actions ($|\mathcal{A}|$) & 6 \\
State dimension ($|\mathcal{S}|$) & 53 \\
\hline
\end{tabular}
\end{table}

\subsubsection{Inference Strategy}
To reduce the inference latency, the proposed DIA should not be activated for every frame. Instead, its activation frequency can be decided based on the specific gameplay scenario. Specifically, since game players' eyes usually fixate on a single object for hundreds of milliseconds \cite{duration}, we set the DIA to make resolution decisions every 250 ms, which is equivalent to every 30 frames at a frame rate of 120 FPS. 

Additionally, the proposed DIA operates in a non-blocking manner. Once a bitrate allocation decision is generated, it remains effective and is continuously applied to subsequent frames until a new decision becomes available. Importantly, the rendering and streaming pipeline never waits for the completion of object identification or policy inference. As a result, the computational overhead in the decision process only affects the timeliness of adaptation, rather than the end-to-end latency. By using an RTX 4090 GPU, the average inference time of the semantic segmentation model is about 90 ms, and the DRL model is just 0.96 ms. Therefore, the total additional latency introduced by DIA is much less than the 250 ms decision interval, allowing a new bitrate allocation to be generated before the next update without blocking the rendering and streaming pipeline. Furthermore, in fully integrated cloud gaming platforms, the object semantic segmentation model can be replaced by direct metadata extraction from the game engine's rendering pipeline (e.g., accessing object IDs), effectively reducing the object identification compute cost to near-zero. 

\subsection{Integration with Production Encoders}
DIA is designed to be compatible with standard commercial cloud gaming architectures. It functions as a bitrate control module that interacts with standard video encoders (e.g., H.264/HEVC) via \textbf{ROI APIs}. Specifically, the resolution decisions made by the DIA agent are mapped to a \textbf{Quantization Parameter (QP) Map} (or Delta-QP values). This QP map is then passed to the encoder (e.g., x265), which executes the compression. This decoupled design ensures that DIA can be deployed without modifying the underlying codec standards.

\section{Evaluation}
In this section, we first introduce the evaluation metrics. Then, we present the baseline approaches followed by the experiment settings. Finally, we present the evaluation results and demonstrate a subjective user study. 
\subsection{Evaluation Metrics}{\label{sec:QoE}}
There are conventional QoE metrics \cite{Pensieve} for evaluating the performance of ABR schemes in video streaming. However, they cannot be directly applied to cloud gaming, since they treat all objects within the same frame equally, which is not suitable for cloud gaming, where users typically focus on a few key objects (e.g., vehicles, pedestrians). Therefore, we propose three new QoE metrics that are more suitable for cloud gaming by replacing the uniform frame-level bitrate utility with a weighted object-level utility from the conventional QoE models \cite{Pensieve}. Each of them is a variant of Eq.~\eqref{eq:QoE} with different bitrate quality functions and configurable weights. Table~\ref{tab:qoe_metrics} shows the specific choices of the bitrate quality functions and weights (some of the weights are selected based on \cite{Pensieve}) that are used in each metric.



\begin{itemize}
  \item $QoE_{lin}$: This metric is proposed to evaluate the impacts of rebuffering time on users' QoE. It directly uses the raw values of each factor in a linear combination. To ensure that the bitrate and rebuffering time are on the same scale during calculation, the weight $\beta$ for rebuffering time should be set significantly greater than those of other factors. Moreover, to guarantee the key objects are assigned a higher bitrate, we set $\omega_k \in (0.5, 1]$.  
  \item $QoE_{log}$: Since researchers claimed that the relationship between video quality and bitrate is non-linear \cite{360}, we adopt a logarithmic function to represent the bitrate quality. 
  \item $QoE_{HD}$: This metric is designed to evaluate if key objects are delivered at high resolutions without introducing high rebuffering time.
\end{itemize}

\begin{table*}[!t]
\centering
\caption{Proposed QoE Metrics}
\label{tab:qoe_metrics}
\renewcommand{\arraystretch}{1.6}
\begin{tabular}{|l|c|c|c|c|c|}
\hline
\textbf{Metric} & \textbf{Bitrate Quality} $Q_{br}(o_k,t)$ & $\alpha$ & $\beta$ & $\gamma$ & $\theta$ \\
\hline
$QoE_{lin}$ &
$Q_{br}(o_k, t) = \omega \cdot br_t(o_k), \quad \text{where } 
\omega =
\begin{cases}
\omega_k, & \text{if } o_k \text{ is a key object} \\
1 - \omega_k, & \text{otherwise}
\end{cases}$&
$1$ &
$430$ &
$1$ &
$1$ \\
\hline
$QoE_{log}$ &
$Q_{br}(o_k, t) = log(\omega \cdot \frac{br_t(o_k)}{br_{min}}), \quad \text{where } 
\omega =
\begin{cases}
\omega_k, & \text{if } o_k \text{ is a key object} \\
1 - \omega_k, & \text{otherwise}
\end{cases}$&
$1$ &
$4.3$ &
$1$ &
$1$ \\
\hline
$QoE_{HD}$ &
$Q_{br}(o_k, t) \rightarrow
\begin{cases}
1, & \text{if } r(o_k) \leq 1280 \times 720 \\
2, & \text{if } r(o_k) = 1920 \times 1080 \\
12, & \text{if } r(o_k) = 2560 \times 1440 \\
15, & \text{if } r(o_k) = 3840 \times 2160
\end{cases}$&
$1$ &
$8$ &
$1$ &
$1$ \\
\hline
\end{tabular}
\end{table*}

More specifically, the proposed metrics preserve the general structure of conventional ABR QoE formulations, including bitrate utility, rebuffering penalty, quality-variation penalty, and latency penalty. The main difference is that the frame-level bitrate utility is replaced by an object-weighted utility so that the delivered quality of gameplay-relevant objects contributes more strongly to the overall QoE. This extension allows the metrics to distinguish two frames with similar average bitrate but different quality allocations across key and non-key objects.

The three metrics characterize complementary aspects of user experience. The linear-utility metric reflects the direct trade-off between delivered quality and transmission cost. The logarithmic-utility metric captures diminishing perceptual returns when bitrate increases at already high quality levels. The HD-oriented metric emphasizes whether important gameplay objects are preserved at sufficiently high resolutions, which is particularly relevant in cloud gaming because the visibility of targets, vehicles, pedestrians, and other task-relevant objects may affect both visual perception and gameplay effectiveness.

Furthermore, to avoid evaluating DIA solely based on the proposed QoE formulations, we also report several conventional Quality of Service (QoS) metrics, including average rebuffering time, average bitrate difference, average bitrate, and average end-to-end latency. All these QoS metrics are calculated at the frame level. In addition, we conduct a subjective user study to directly assess users’ perceived visual quality under different schemes. The subjective study provides an independent perceptual assessment and complements the objective QoE and QoS evaluations.

\subsection{Baseline Approaches}
To comprehensively evaluate the performance of the proposed DIA scheme, we compare it against two categories of baseline approaches: conventional chunk-based ABR schemes (BBA, BOLA, RobustMPC, and Pensieve) and ROI-based cloud gaming ABR schemes (DeepGame and T-Gaming).
\begin{itemize}
  \item \textbf{BOLA}: This baseline approach selects the bitrate by solving the Lyapunov optimization problem \cite{bola}.
  \item \textbf{RobustMPC}: This baseline approach leverages an MPC model to predict the throughput and buffer occupancy, and then selects a suitable bitrate to maximize the QoE \cite{RobustMPC}. 
  \item \textbf{Pensieve}: This baseline approach is a DRL-based ABR scheme that learns the optimal bitrate selection policy based on current observations, and it outperforms conventional SOTA approaches \cite{Pensieve}.
  \item \textbf{DeepGame}: DeepGame utilizes LSTM to predict the ROI, and prioritizes high bitrate to the predicted ROI \cite{deepgame}.
  \item \textbf{T-Gaming}: T-Gaming identifies the saliency map as the ROI and allocates high bitrate to these visually salient areas \cite{T-gaming}.
  \item \textbf{Nebula}: Nebula adaptively controls the video encoder parameter and the amount of frame-level redundancy based on the measured network conditions \cite{Nebula}.
  \item \textbf{MetaABR}: MetaABR is a meta-reinforcement learning-based ABR scheme that jointly trains multiple learning tasks with a shared meta-critic to provide transferable knowledge for bitrate selection across different network environments and QoE objectives \cite{MetaABR}. 
  
\end{itemize}

\subsection{Simulation Settings}
To comprehensively evaluate the performance of the proposed DIA scheme under different network conditions, we developed a custom simulator in Python. This simulator models the dynamic network conditions and user mobility patterns in various wireless network scenarios. In our experiments, we consider three typical wireless scenarios: 5G, 4G, and WiFi. All scenarios are simulated in the same geographical area with the same number of users, but with different user mobility patterns and network parameters. 

The total number of users in the simulation is 160, and they can enter or leave this area during the simulation. The initial number and position of users are randomly generated. The simulation duration is set to 300 seconds, which is sufficient to capture the dynamic network conditions and user mobility patterns. We perform 10 runs for each scenario and calculate the average over these 10 runs to obtain the simulation results. We adopt representative values of bandwidth for different network scenarios based on existing works \cite{audio, A2BR} and practical considerations. To capture the stochastic nature of wireless channels, the simulator implements a \textbf{distance-dependent path loss model}, where the signal quality decays logarithmically with the distance between the user and the base station. Interference and background traffic are modeled by adding a Gaussian perturbation to the effective available bandwidth in each time slot. Furthermore, a \textbf{velocity-dependent degradation factor} is introduced to model the impact of Doppler shifts and fast fading in high-mobility scenarios (e.g., 5G/4G driving). Handover behavior is modeled by the dynamic fluctuation of available bandwidth as users move across the coverage area. Then, the bandwidth value $bw_t(u_n)$ for user $u_n$ at time slot $t$ is generated by the simulator after considering the effects of user mobility, interference, and background traffic, and the duration of each time slot $t$ is set to 1 second. Within each time slot, DIA can update its object-level decisions multiple times according to the update frequency described in Section IV-D. The details of the network parameters are as follows:

\begin{itemize}
  \item \textbf{5G Network:} The users' speed ranges from 0 to 30 km/h, and the maximum available bandwidth for a user is set to 300 Mbps, which is the typical median throughput of commercial 5G networks observed in an urban environment \cite{5G}. 
  \item \textbf{4G Network:} The range of users' speed is set to be the same as 5G networks, i.e., from 0 to 30 km/h, and the maximum available bandwidth for a user is set to 100 Mbps, consistent with the typical performance of LTE-Advanced networks \cite{5G}.
  \item \textbf{WiFi Network:} Since most WiFi users are stationary, we set users' speed between 0 km/h and 5 km/h, and the maximum available bandwidth for a user is the same as in the 5G networks (300 Mbps), reflecting the practical bandwidth of modern WiFi standards \cite{wifi}.
\end{itemize}


\subsection{Evaluation Results}

\textbf{Key-object Identification Accuracy.}
Before evaluating the performance of DIA, we further examine whether DIA can correctly identify key objects. Since most baseline ABR schemes operate at the frame, chunk, or stream level and do not explicitly output key-object labels, this evaluation is conducted as a standalone validation of DIA. Using the human-annotated ground-truth labels with multi-annotator aggregation, DIA achieves an average key-object identification accuracy of 90.58\% over 10 evaluation runs. This result indicates that DIA can effectively identify perceptually important objects in the evaluated gameplay scenes.

\begin{figure}[!htp]
\centering
\subfloat[5G Network]{\includegraphics[width=3.2in]{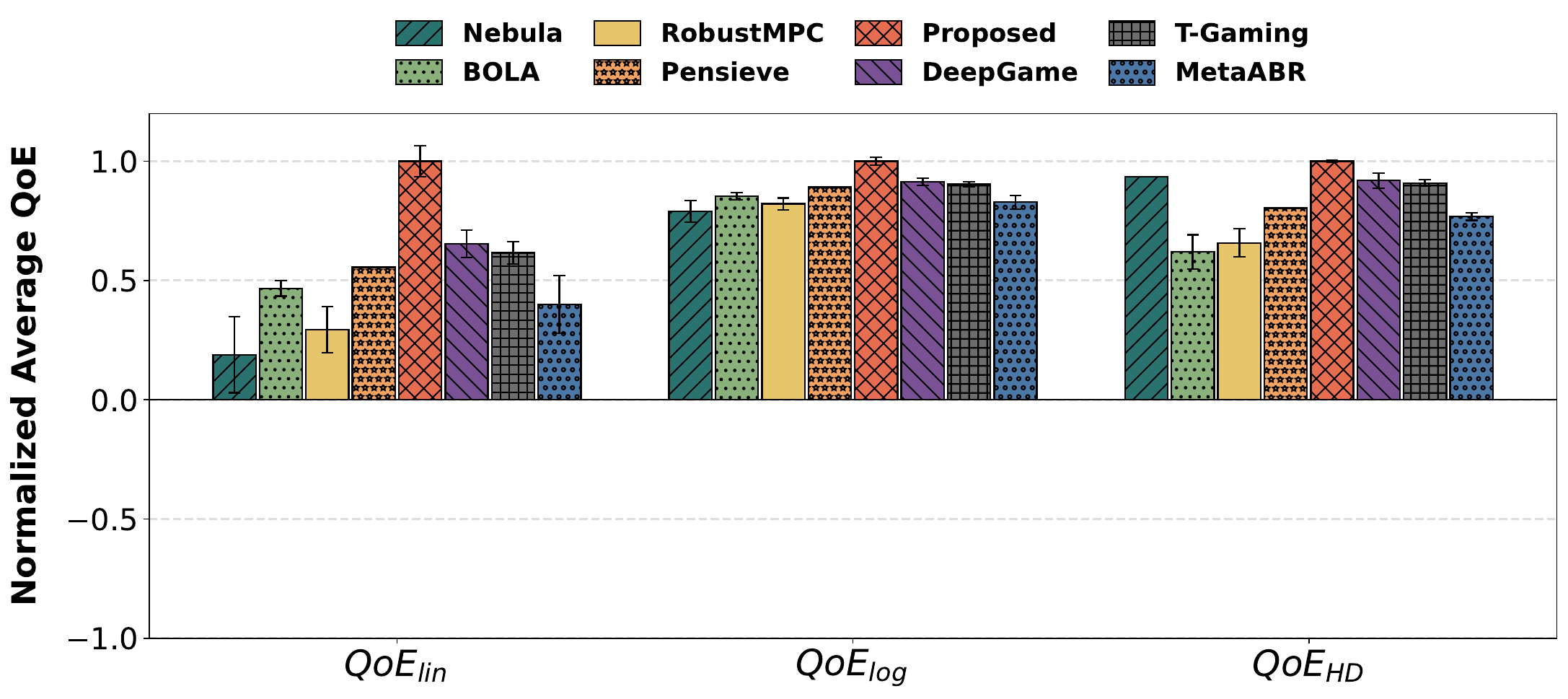}\label{fig:qoe_5g}}
\hfil
\subfloat[4G Network]{\includegraphics[width=3.2in]{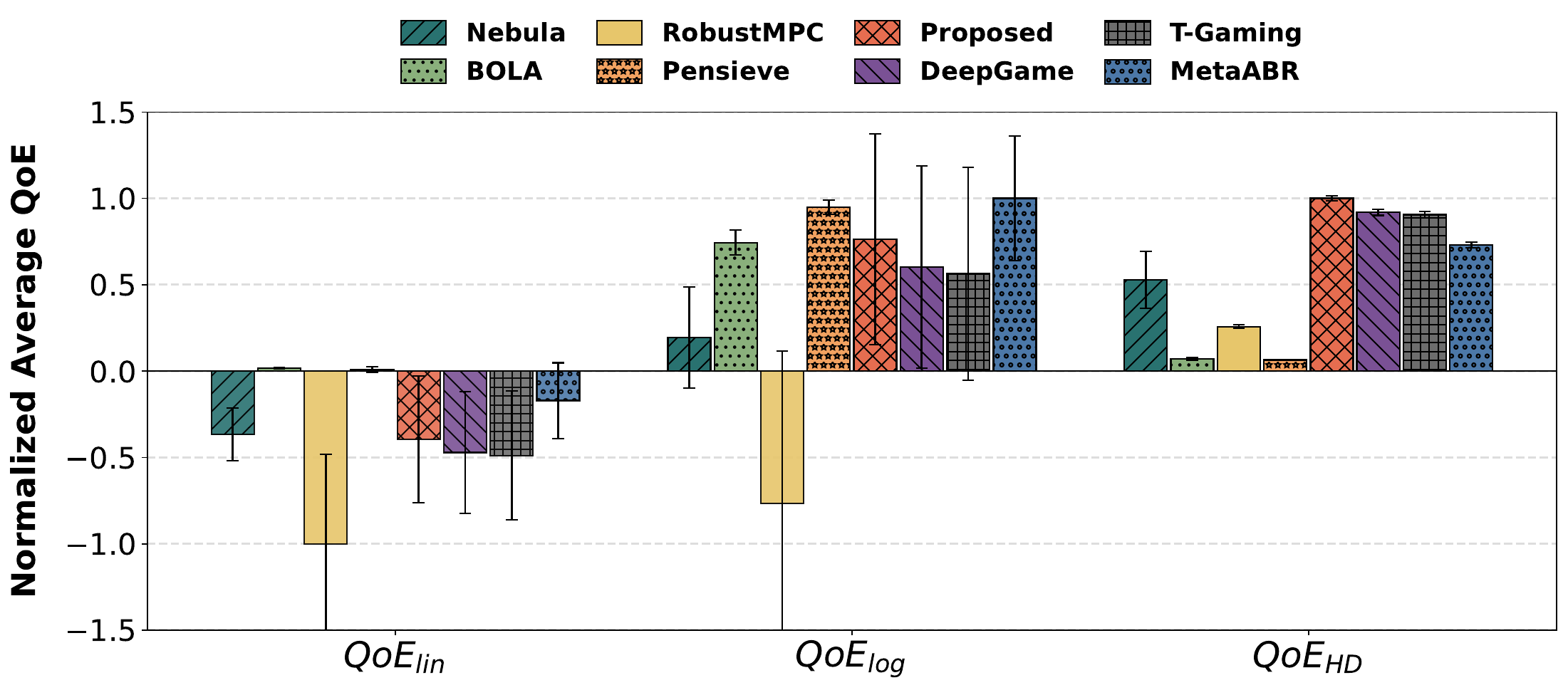}\label{fig:qoe_4g}}
\\
\subfloat[WiFi Network]{\includegraphics[width=3.2in]{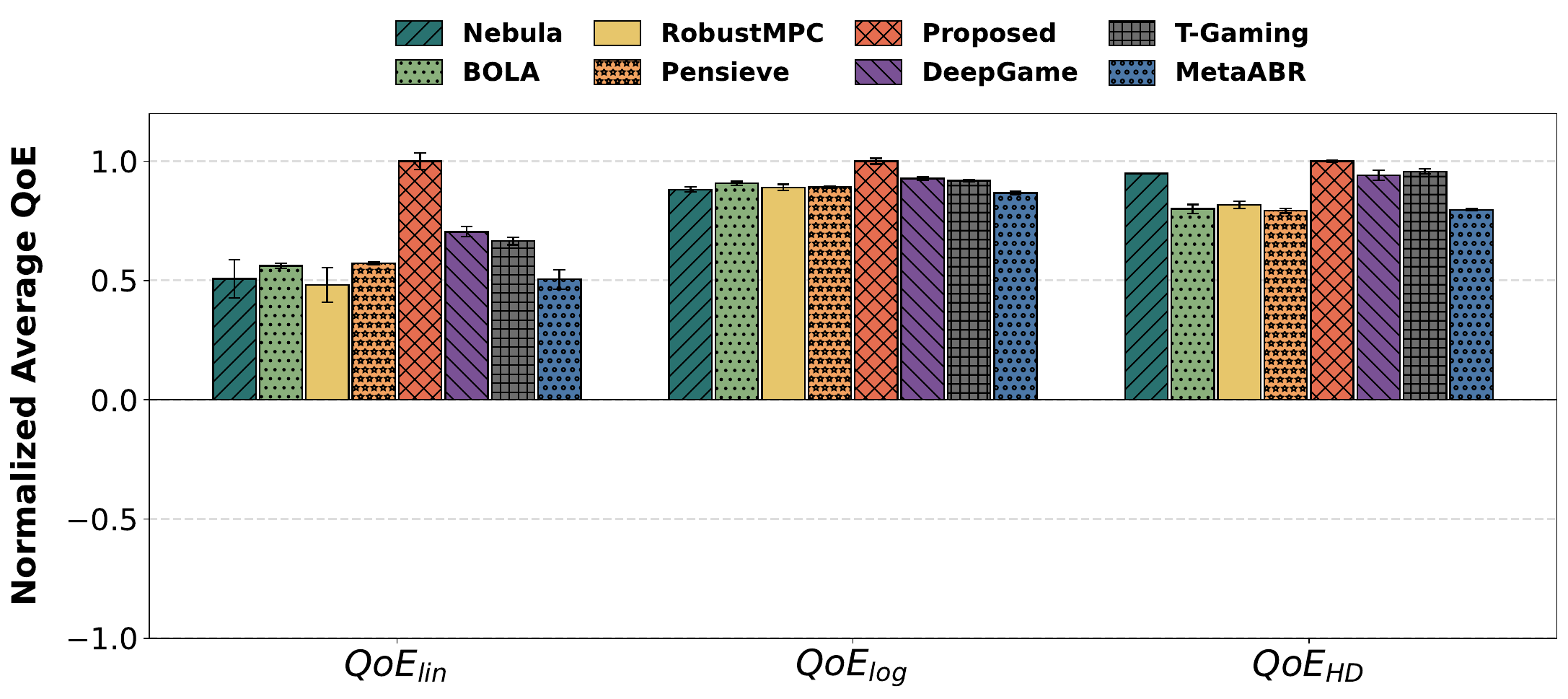}\label{fig:qoe_wifi}}
\caption{QoE comparisons of the proposed DIA and baseline approaches under different network scenarios. Error bars span $\pm$ one standard deviation from the average.}
\label{fig:qoe_networks}
\end{figure}

Fig.~\ref{fig:qoe_networks} shows the normalized average QoE values of the proposed DIA and baseline approaches under different network scenarios. Fig.~\ref{fig:qoe_networks}\subref{fig:qoe_5g} and Fig.~\ref{fig:qoe_networks}\subref{fig:qoe_wifi} illustrate that the proposed DIA outperforms all baseline approaches in terms of all QoE metrics under 5G and WiFi networks. Notably, ROI-based approaches (DeepGame and T-Gaming) achieve higher values than conventional schemes (e.g., RobustMPC, BOLA, and Pensieve) due to their ability to prioritize high bitrates for key objects. However, since they cannot accurately identify key objects (DeepGame relies on object motion history to determine key objects, and T-Gaming only utilizes saliency maps to identify key objects), their performance is still inferior to DIA. As shown in Fig.~\ref{fig:qoe_networks}\subref{fig:qoe_5g}, once the users' mobility increases, the performance difference between DIA and the baseline approaches becomes more significant in terms of $QoE_{log}$ and $QoE_{HD}$. The reason behind this is that the available bandwidth becomes more constrained with the increased mobility, which negatively affects the performance of the baseline approaches. In contrast, the proposed DIA can prioritize high resolutions for key objects in cases where bandwidth is limited, resulting in better $QoE_{log}$ and $QoE_{HD}$.


\begin{figure*}[!htp]
\centering
\subfloat[$QoE_{lin}$ ]{\includegraphics[width=2.1in]{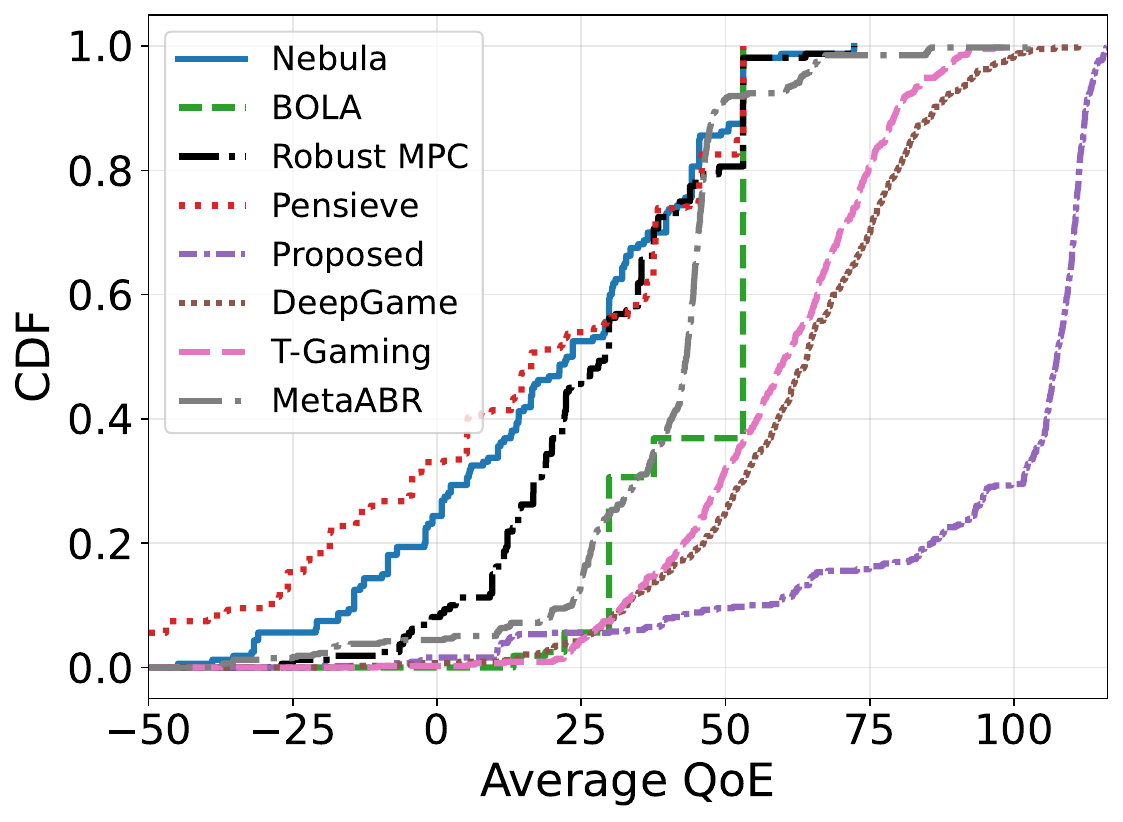}\label{fig:qoe_lin_cdf}}
\hfil
\subfloat[$QoE_{log}$ ]{\includegraphics[width=2.1in]{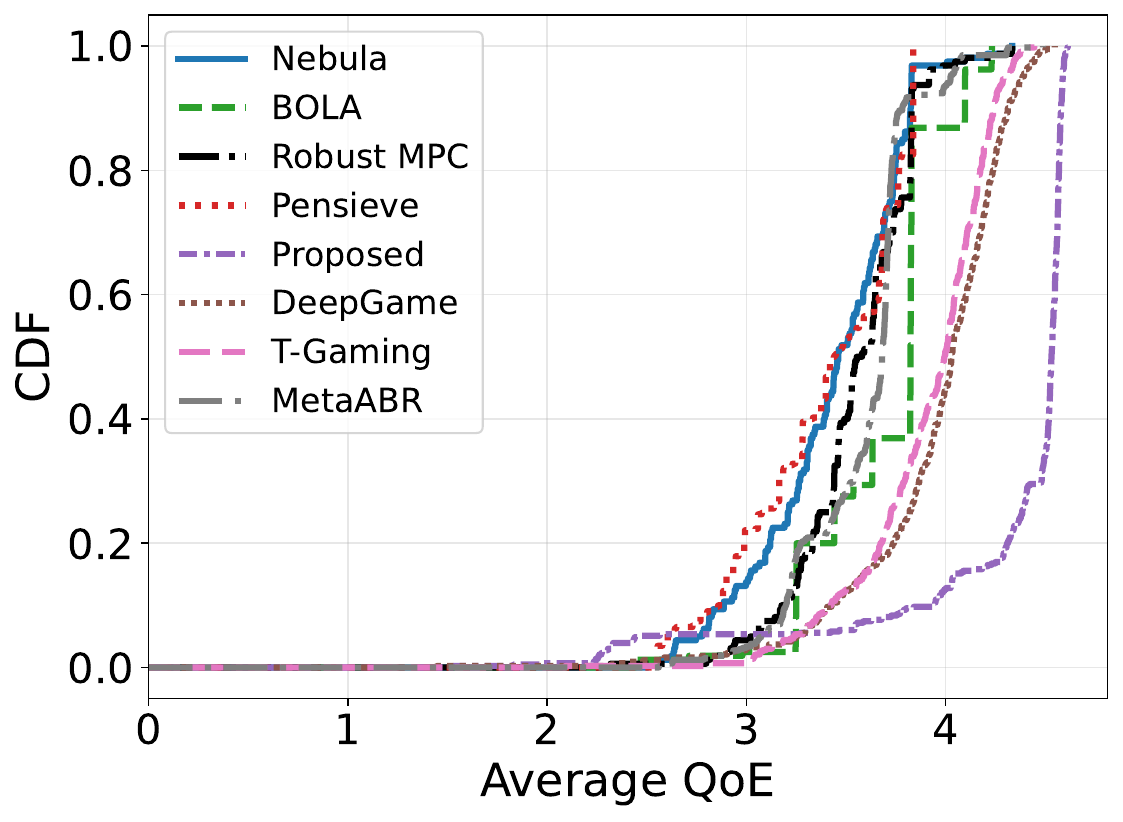}\label{fig:qoe_log_cdf}}
\hfil
\subfloat[$QoE_{HD}$ ]{\includegraphics[width=2.1in]{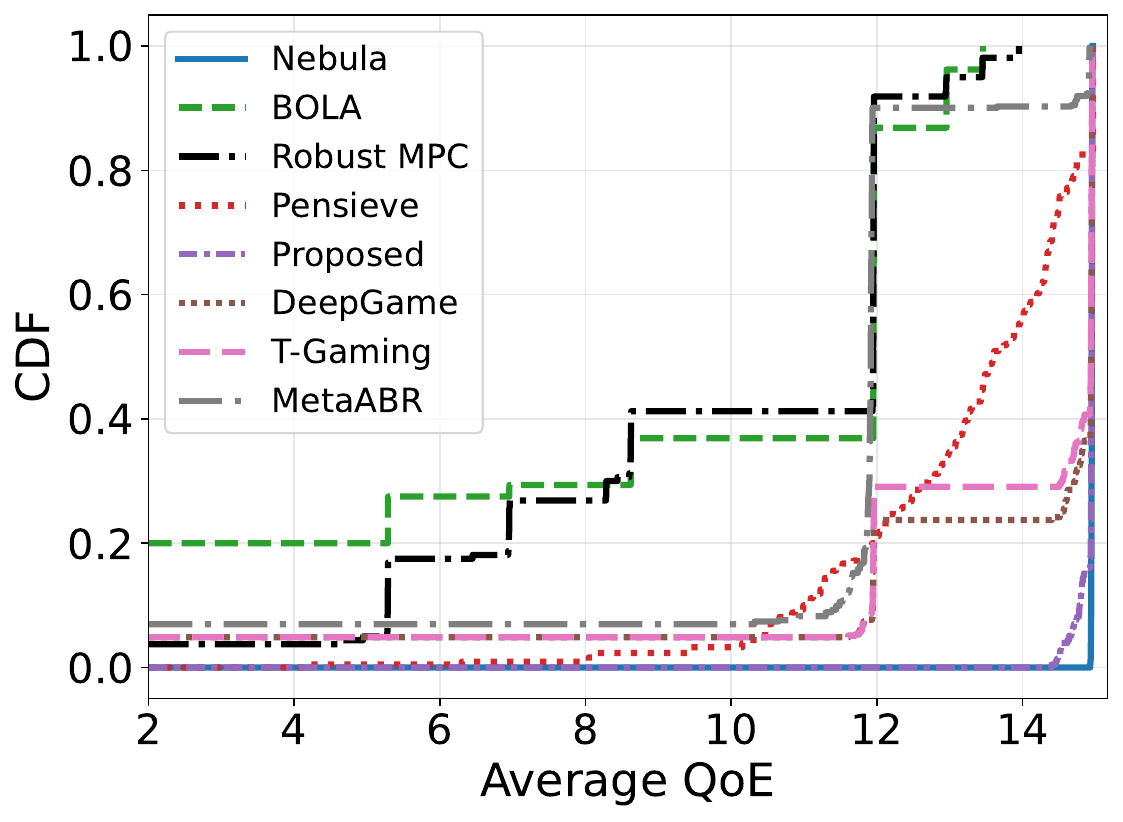}\label{fig:qoe_hd_cdf}}
\caption{Comparisons of the proposed DIA with baseline approaches in terms of the proposed QoE metrics under 5G networks.}
\label{fig:qoe_cdf}
\end{figure*}

\begin{figure*}[!htp]
\centering
\subfloat[$QoE_{lin}$ ]{\includegraphics[width=2.1in]{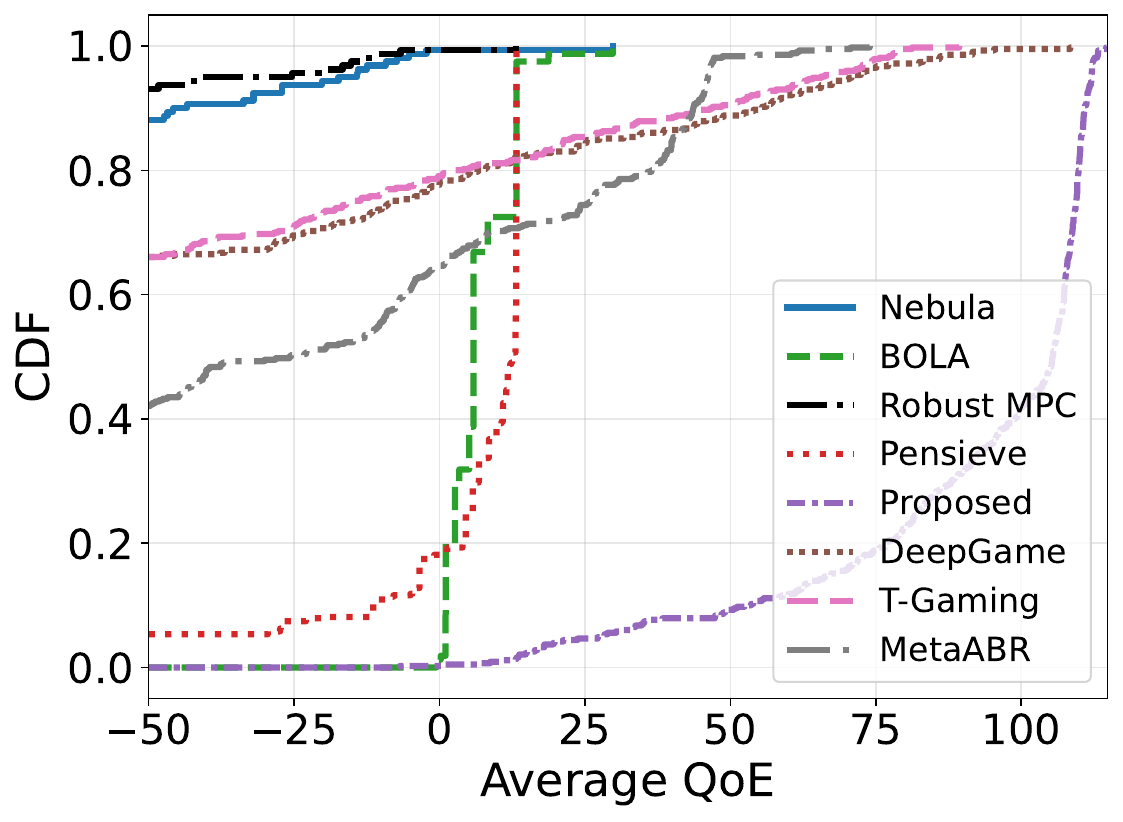}\label{fig:4G_qoe_normal_cdf}}
\hfil
\subfloat[$QoE_{log}$ ]{\includegraphics[width=2.1in]{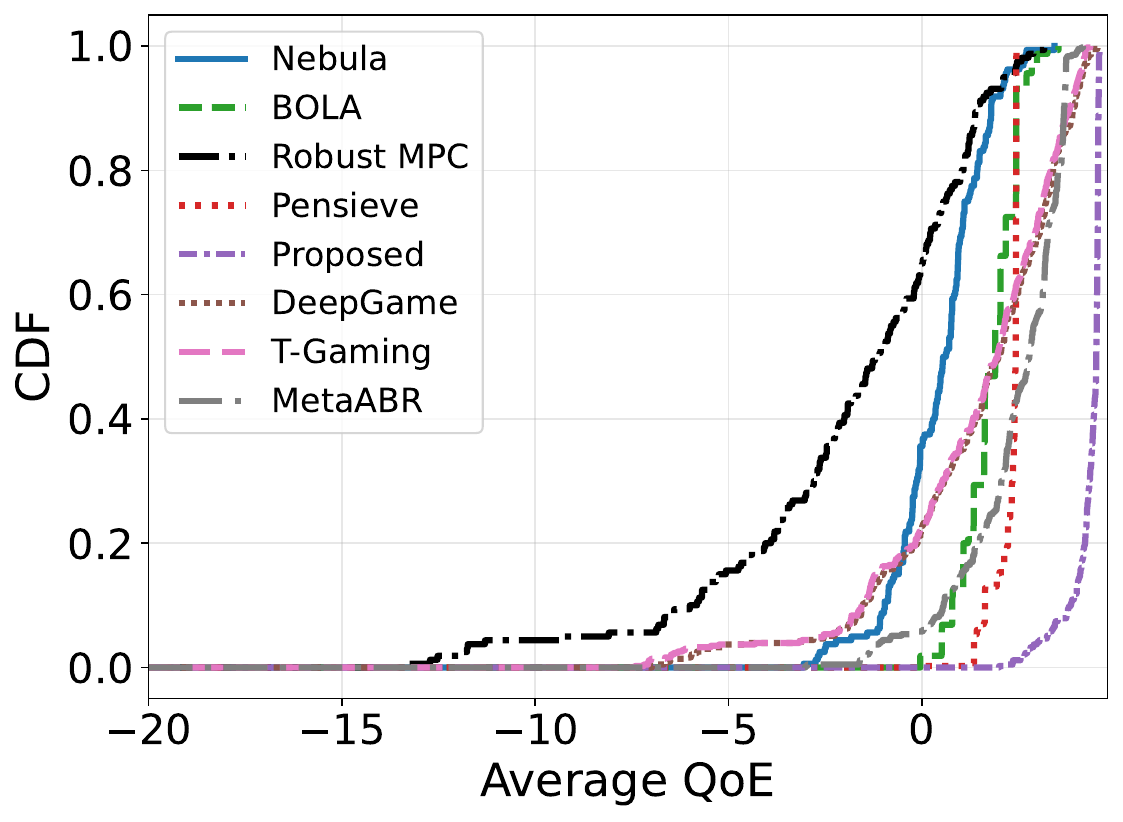}\label{fig:4G_qoe_log_cdf}}
\hfil
\subfloat[$QoE_{HD}$ ]{\includegraphics[width=2.1in]{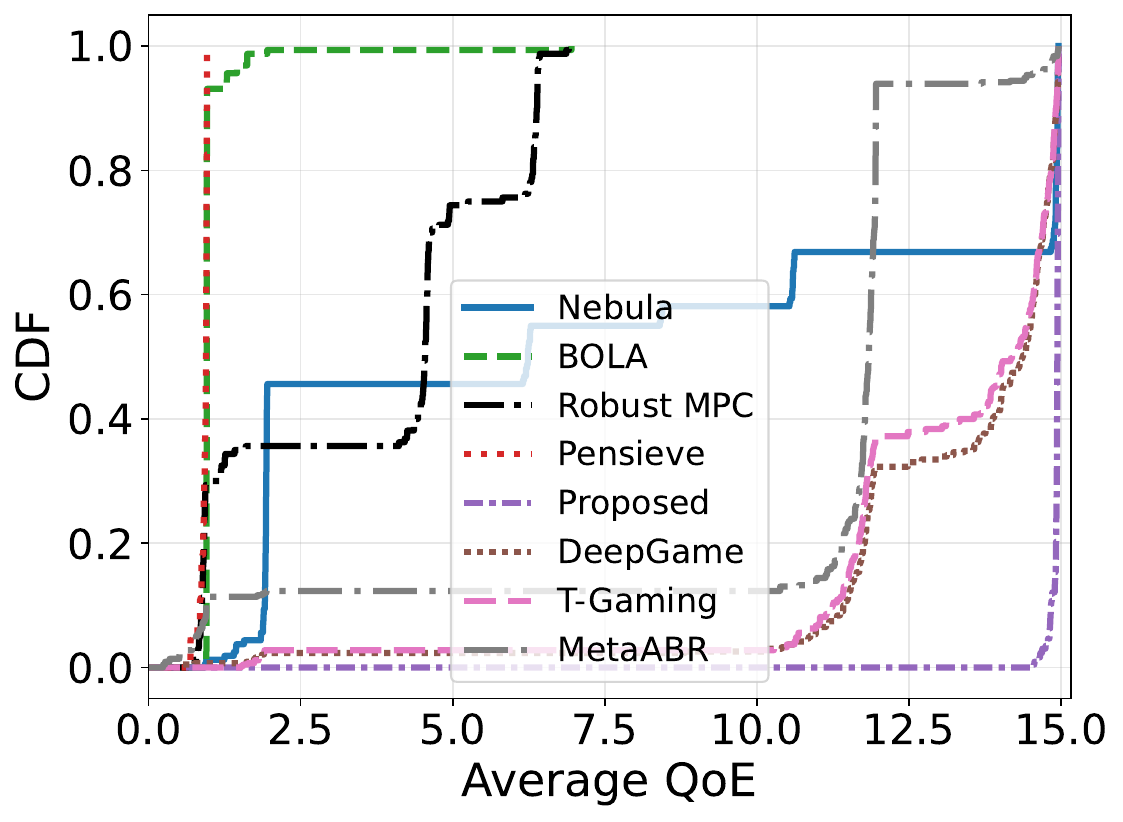}\label{fig:4G_qoe_hd_cdf}}
\caption{Comparisons of the proposed DIA with baseline approaches in terms of the proposed QoE metrics under 4G networks.}
\label{fig:4G_qoe_cdf}
\end{figure*}

\begin{figure*}[!htp]
\centering
\subfloat[$QoE_{lin}$ ]{\includegraphics[width=2.1in]{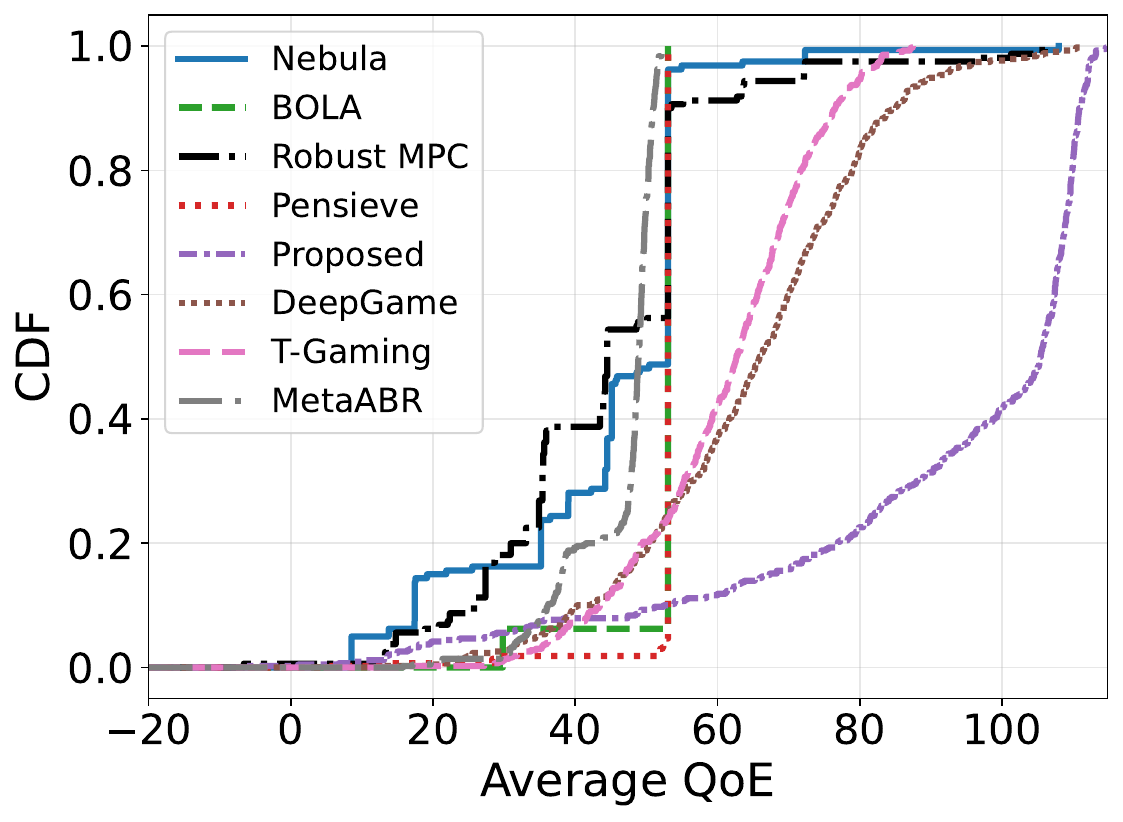}\label{fig:wifi_qoe_lin_cdf}}
\hfil
\subfloat[$QoE_{log}$ ]{\includegraphics[width=2.1in]{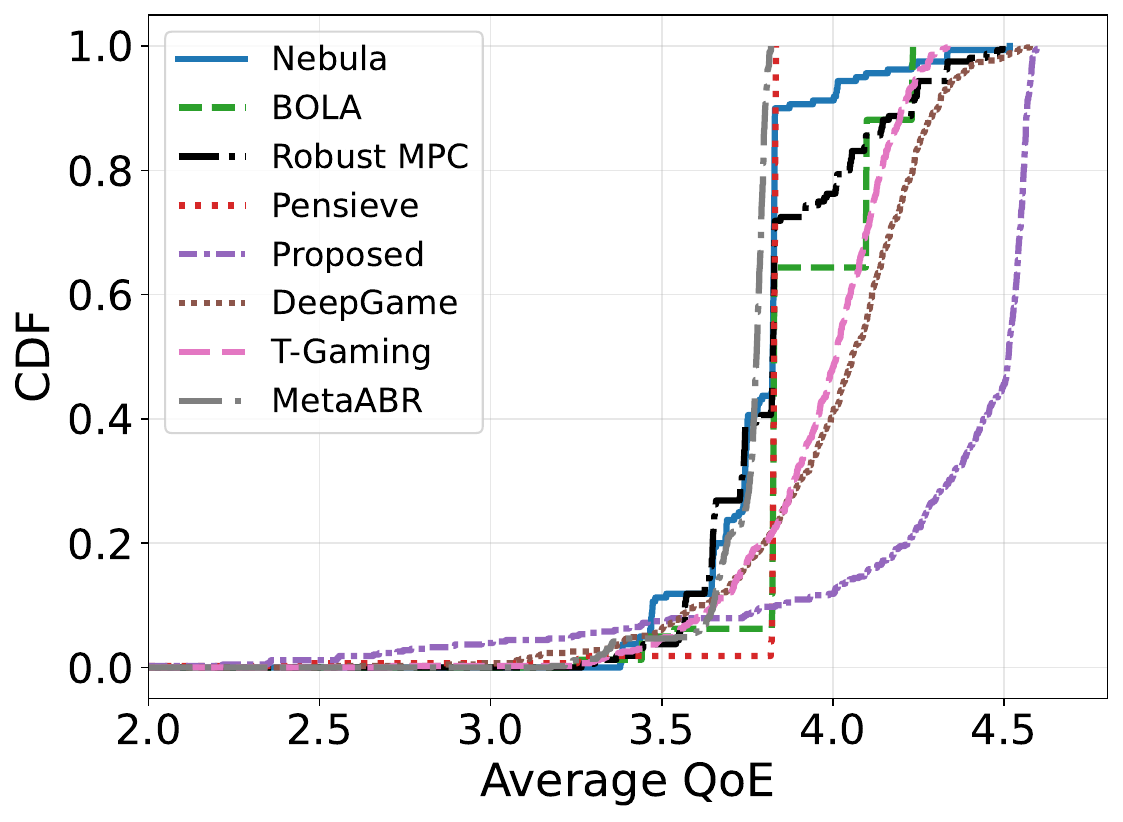}\label{fig:wifi_qoe_log_cdf}}
\hfil
\subfloat[$QoE_{HD}$ ]{\includegraphics[width=2.1in]{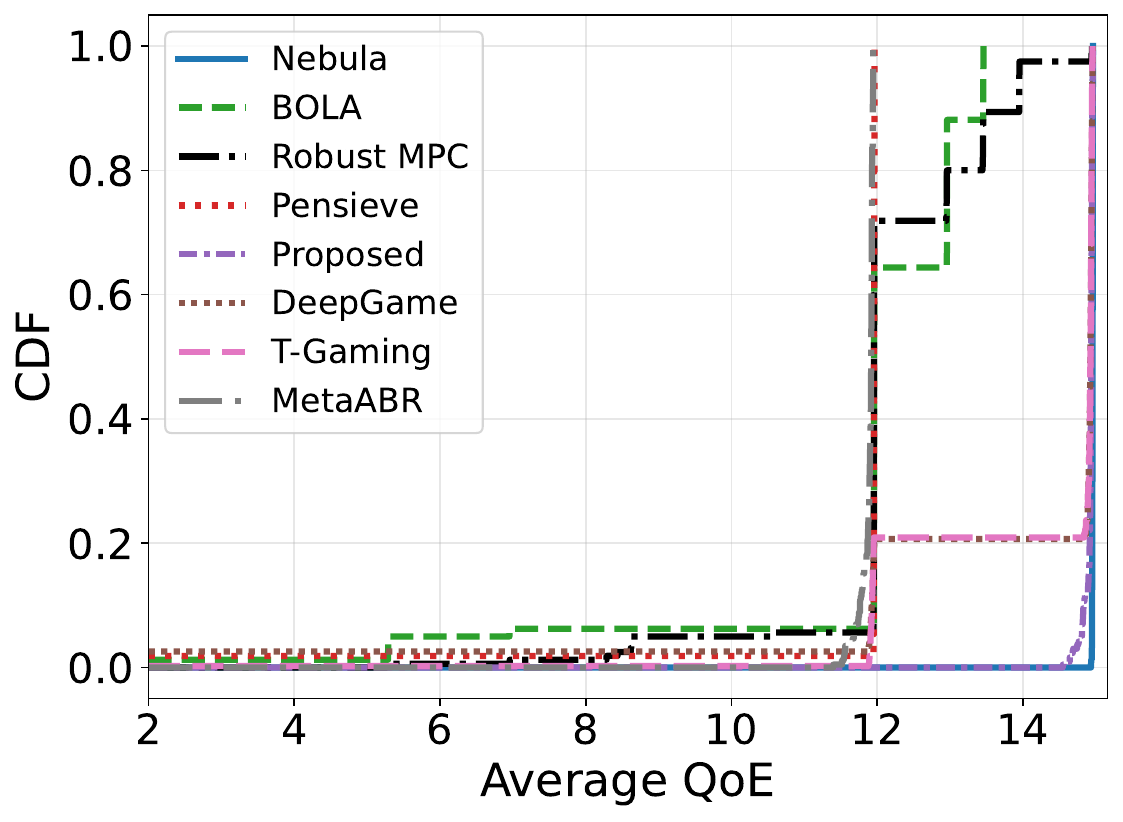}\label{fig:wifi_qoe_hd_cdf}}
\caption{Comparisons of the proposed DIA with baseline approaches in terms of the proposed QoE metrics under WiFi networks.}
\label{fig:wifi_qoe_cdf}
\end{figure*}

Fig.~\ref{fig:qoe_networks}\subref{fig:qoe_4g} demonstrates the performance comparison under 4G networks, which can be treated as a limited bandwidth scenario. We observe that only BOLA and Pensieve achieve positive normalized QoE values in terms of $QoE_{lin}$. However, this comes at the expense of reduced key objects' resolution, as shown by their low $QoE_{HD}$ among all approaches. In contrast, the performance of DeepGame and T-Gaming is slightly lower than DIA. This is also because they cannot accurately identify key objects, leading to suboptimal bitrate allocation. Additionally, we can see that the error bars for RobustMPC and Pensieve are much larger than those for DIA, indicating that these baselines do not adapt well to the dynamic network conditions when bandwidth is limited. In contrast, DIA has a much smaller error bar, demonstrating its ability to adapt more effectively. MetaABR outperforms DIA in terms of $QoE_{lin}$ and $QoE_{log}$, but it perform much worse than DIA in terms of $QoE_{HD}$. This is because it cannot explicitly prioritize high resolutions for key objects, which is crucial for achieving high $QoE_{HD}$ under bandwidth-constrained scenarios.

Fig.~\ref{fig:qoe_cdf} to Fig.~\ref{fig:wifi_qoe_cdf} provide a more detailed comparison of the proposed DIA scheme with baseline approaches by displaying the Cumulative Distribution Function (CDF) comparisons under 5G, 4G, and WiFi networks, respectively. Overall, the proposed DIA significantly outperforms baseline approaches across all metrics and network settings.

\begin{figure*}[htp]
\centering
\subfloat[5G Network]{\includegraphics[width=6.2in]{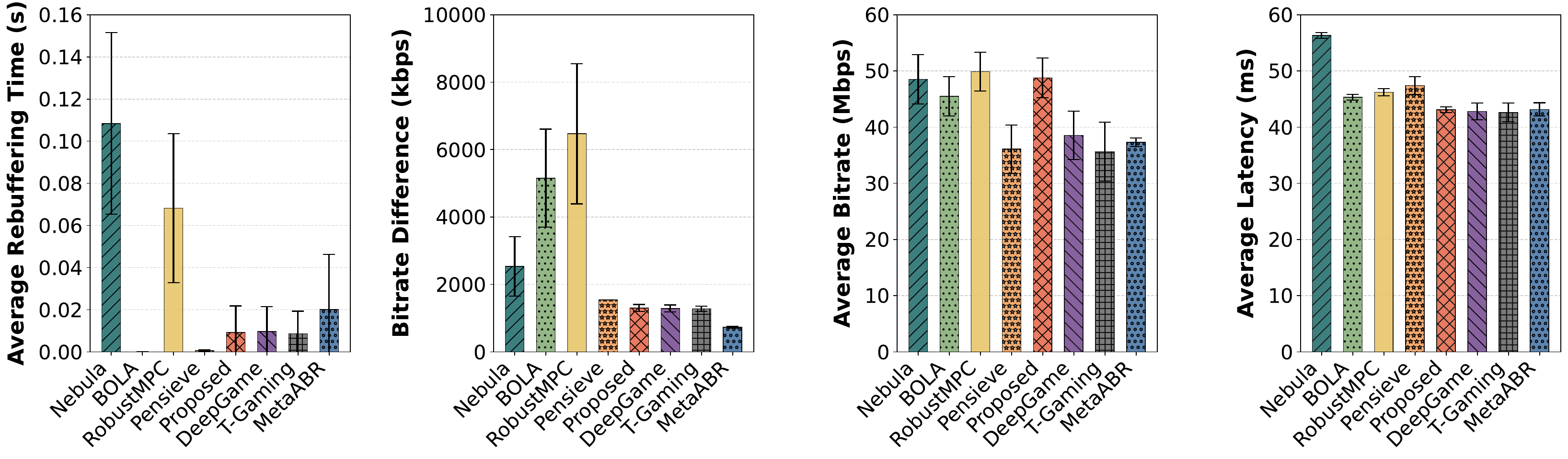}\label{fig:qos_5g}}
\\
\subfloat[4G Network]{\includegraphics[width=6.2in]{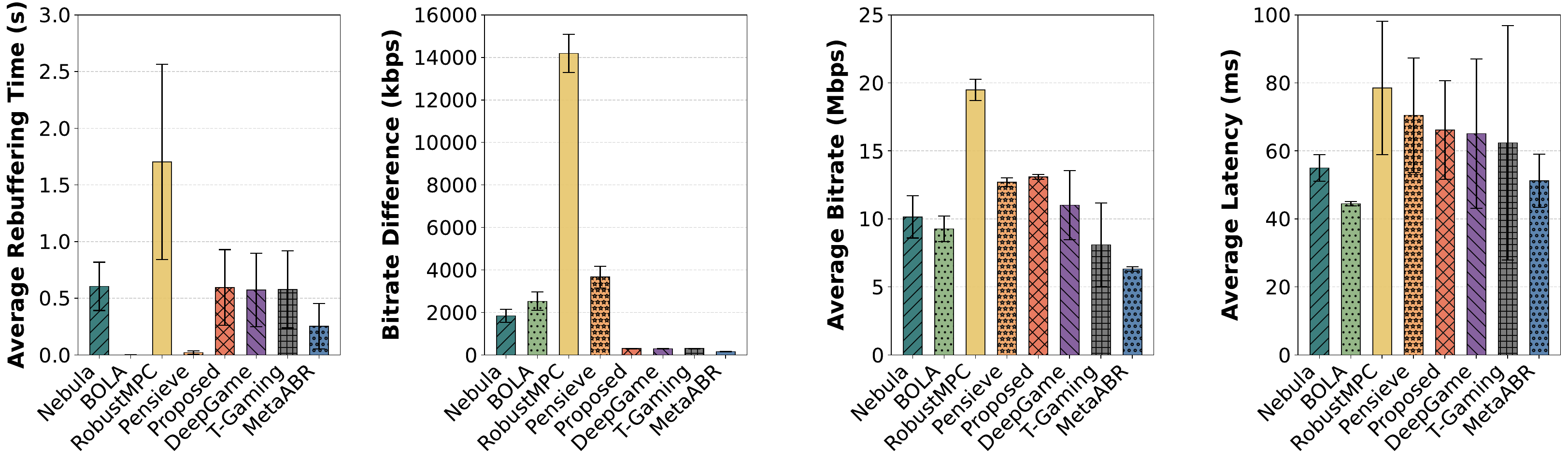}\label{fig:qos_4g}}
\\
\subfloat[WiFi Network]{\includegraphics[width=6.2in]{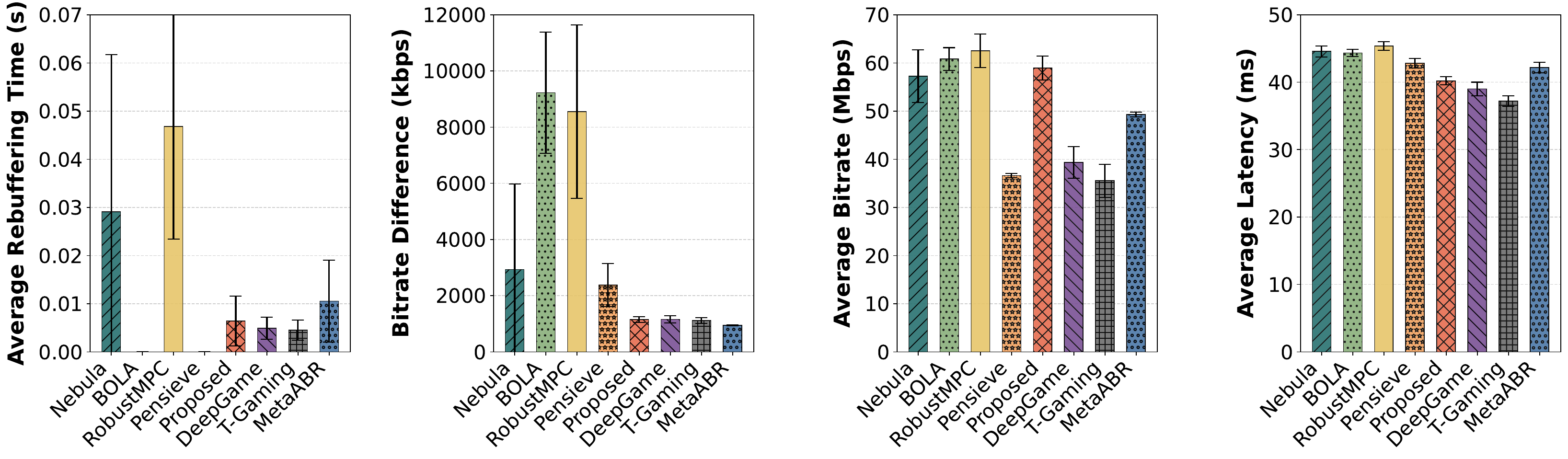}\label{fig:qos_wifi}}
\caption{QoS comparison of the proposed DIA and baseline approaches under different network scenarios. Error bars span $\pm$ one standard deviation from the average.}
\label{fig:qos_networks}
\end{figure*}

For example, in high bandwidth scenarios like 5G and WiFi networks (Fig.~\ref{fig:qoe_cdf} and Fig.~\ref{fig:wifi_qoe_cdf}), by using DIA, more than 80\% of users' $QoE_{lin}$ values are higher than 80, while more than 90\% of baseline approach users' $QoE_{lin}$ values are less than 60; all users' $QoE_{HD}$ values are higher than 14 by using DIA, while only a small amount of baseline approach users' $QoE_{HD}$ values are higher than 14. This is because all baseline approaches work at the frame level and cannot guarantee key objects' quality. To achieve the same quality for key objects as DIA, they have to allocate a higher bitrate to the entire frame, which requires more bandwidth and results in higher rebuffering time and end-to-end latency.

In the bandwidth-constrained scenario, such as 4G networks (Fig.~\ref{fig:4G_qoe_cdf}), DIA still maintains a clear performance advantage. Obviously, except BOLA, Pensieve, and DIA, all approaches suffer negative values in terms of $QoE_{lin}$ and $QoE_{log}$. Although BOLA and Pensieve avoid negative QoE values in $QoE_{lin}$ and $QoE_{log}$, most users experience QoE values within a narrow range (e.g., 0–10 in terms of $QoE_{lin}$ for BOLA, and about 0-15 $QoE_{lin}$ values for Pensieve), far below the high QoE range achieved by DIA, which reaches up to 100 in terms of $QoE_{lin}$. Meanwhile, DeepGame and T-Gaming achieve a similar performance in all QoE metrics, but due to their inability to accurately identify key objects, their performance is still inferior to DIA.

Fig.~\ref{fig:qos_networks} shows the QoS comparisons of the proposed DIA and baseline approaches under different network scenarios. These QoS metrics (including average rebuffering time, average bitrate difference, average bitrate, and average latency) provide deeper insights into the underlying reasons for the observed QoE performance, as these QoS metrics have strong impacts on the calculations of the QoE metrics. 


For instance, one observation is that high bitrate does not always lead to better QoE performance. From Fig.~\ref{fig:qos_networks}\subref{fig:qos_5g} to \subref{fig:qos_wifi}, we can see that RobustMPC achieves significantly high average bitrate regardless of the network scenarios. However, its QoE performance is significantly worst than most of the baseline approaches. The reason for this is that RobustMPC tends to select a relatively high bitrate based on estimated bandwidth. However, in highly dynamic wireless cloud gaming scenarios, bandwidth prediction errors can easily lead to overly aggressive bitrate decisions, which results in high rebuffering time, latency and bitrate difference. Consequently, its QoE performance is significantly degraded.

Fig.~\ref{fig:qos_networks}\subref{fig:qos_4g} provides further insights into the QoS performance under 4G networks. We observe that DIA achieves a better balance among bitrate, rebuffering time, bitrate variation, and latency. Although all learning-based approaches (Pensieve, MetaABR, DeepGame, and T-Gaming) have lower latency and rebuffering time than DIA, this comes at the cost of reducing the bitrate of gameplay frames, which leads to a degradation of the QoE performance. In contrast, DIA can maintain a relatively high bitrate while keeping the rebuffering time and latency at a low level, which results in superior QoE performance.


As shown in Fig.~\ref{fig:qos_networks}\subref{fig:qos_wifi}, all approaches achieve a similar high average bitrate and low latency under WiFi networks, since most of the WiFi users are stationary and the available bandwidth is sufficient. Notably, DIA obtains the lowest bitrate difference and the second lowest average rebuffering time, which indicates that it can maintain a more stable bitrate over time compared to other approaches. This is crucial for cloud gaming, as it reduces the risk of sudden quality drops and enhances the overall user experience.

To further examine the consistency between the proposed QoE metrics and user-perceived quality, we conducted a subjective MOS study and compared the resulting MOS values with the three QoE formulations. The relative performance trend obtained from all three QoE metrics is consistent with that observed in the subjective evaluation. Detailed results and discussions are provided in the Appendix.
\subsection{Subjective User Study}

To further examine whether the proposed object-aware QoE design is consistent with real user perception, we conducted a subjective user study using gameplay clips from different game genres, including \textit{Black Myth: Wukong}, \textit{Spider-Man 2}, \textit{The Witcher 3}, \textit{CS:GO}, and \textit{Cyberpunk 2077}. Thirty-four participants were invited to rate the processed gameplay clips using a 1--5 opinion score. The detailed study protocol, participant information, rating procedure, and statistical analysis are provided in the Appendix. 

The subjective results lead to two main observations. First, DIA achieves higher user scores (4.16 on average) than other baseline approaches (Pensieve: 2.49 on average, T-Gaming: 3.74 on average, MetaABR: 2.86 on average) across the evaluated gameplay clips. This indicates that improving the visual quality of key objects is consistent with user-perceived QoE improvement and supports the effectiveness of the proposed object-aware QoE design. The three proposed QoE metrics show the same performance trend as the subjective MOS results, with detailed comparisons provided in the Appendix.

Together with the proposed QoE and conventional QoS results, the subjective user study results demonstrate that DIA can effectively improve the user-perceived QoE in cloud gaming scenarios, especially under bandwidth-constrained conditions. The object-aware bitrate allocation strategy allows DIA to prioritize key objects, leading to a more immersive and satisfying gaming experience for users.

Second, the object update frequency has a significant overall effect on the subjective scores. Increasing the frequency from 2 Hz to 4 Hz improves the average MOS from 3.70 to 4.05, and this difference is statistically significant ($p<0.001$). Increasing the frequency further from 4 Hz to 8 Hz leads to only a small MOS improvement from 4.05 to 4.17, which is not statistically significant ($p=0.096$). Hence, 4 Hz is adopted as the default setting because it provides a significant perceptual improvement over 2 Hz while avoiding the substantially higher inference workload associated with 8 Hz.

\subsection{Limitations}

First, although DIA is evaluated on multiple types of games, its training data is mainly collected from GTA V (open-world game scenes). Therefore, the learned object-importance pattern may not directly generalize to game genres with distinct key objects. For example, in sports games such as FIFA or NBA 2K, users' attention is often concentrated on the ball, which is not sufficiently represented in GTA V like training data. Applying DIA to such genres may require additional data collection and model fine-tuning. Second, the current subjective user study is based on processed gameplay clips to isolate the perceptual impact of object-aware bitrate allocation and update frequency. Further deployment-level validation with real-time cloud gaming systems and online encoder adaptation will be considered in our future work.

\section{Conclusion and Future Work}
In this paper, we proposed DIA, a novel DRL-based ABR scheme that can dynamically adjust the resolution of cloud gaming gameplay transmission at the object-level. Specifically, DIA leverages an off-the-shelf semantic segmentation model to achieve object-level bitrate adjustments. The designed transformer-based feature extractor and interest-aware reward function enable DIA to identify key objects in the gameplay frames based on the users' in-game movement speed, semantic labels, and network conditions. It can guarantee the transmission of key objects at a higher resolution, while selectively lowering the resolution of non-key objects to reduce the overall frame bitrate when the bandwidth is limited. To better evaluate the QoE of ABR schemes for cloud gaming, we also proposed three new QoE metrics, which are able to assess the object-level resolution quality, bitrate stability, and overall user-perceived experience. Extensive experiments were conducted, and the results show that DIA outperforms existing SOTA ABR approaches in terms of the proposed QoE metrics, under different network scenarios, including 5G, 4G, and WiFi networks.

For future work, we plan to further evaluate DIA on a broader range of cloud games with different visual styles and interaction dynamics. While the current model is trained mainly on GTA V gameplay, extending DIA to games with domain specific content, such as sports games with ball-centric attention patterns, may require fine tuning or transfer learning. In addition, although the subjective user study provides initial evidence that the proposed object-aware QoE design is consistent with user perceived quality, a full interactive implementation on real-time cloud gaming platforms, with dynamic encoder adaptation and real-time user control remains an important direction for future work.

\bibliographystyle{IEEEtran}

\bibliography{reference}

\begin{IEEEbiography}
[{\includegraphics[width=1in,height=1.25in,clip,keepaspectratio]{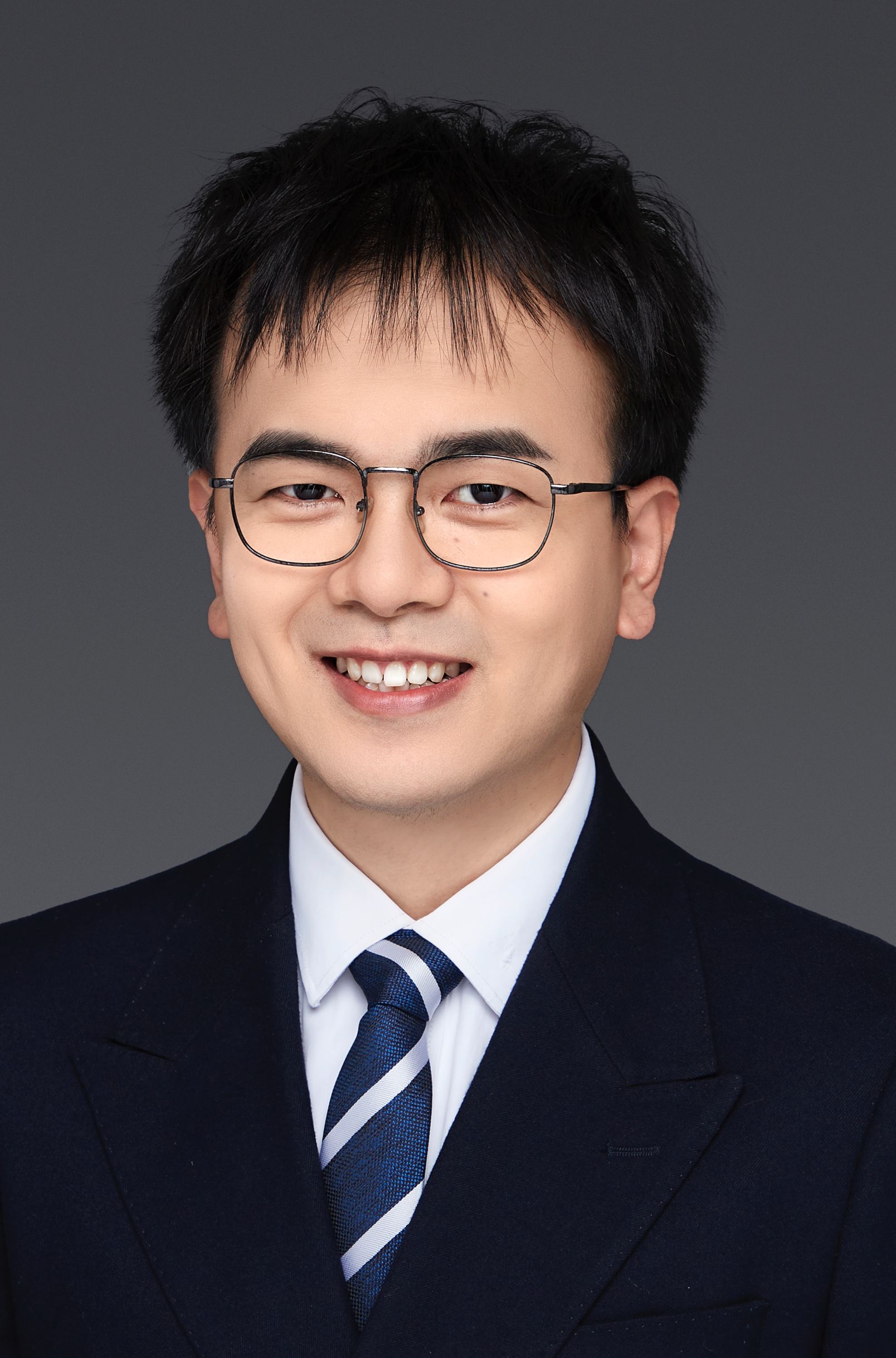}}]{Zhe Zhang}
(Senior Member, IEEE) received his Ph.D. degree in Electrical and Computer Engineering from Carleton University, Ottawa, Canada, in 2019. Currently, he is an Associate Professor with the School of Communications and Information Engineering, Nanjing University of Posts and Telecommunications (NJUPT), Nanjing, Jiangsu, China. Before he joined NJUPT, he was a research engineer at Huawei Ottawa Research Center, Ottawa, Canada. His research interests include multimedia networking, intelligent networks, the metaverse, and in-network caching. He also serves as an Associate Editor of ACM Transactions on Multimedia Computing, Communications, and Applications.
\end{IEEEbiography}
 \vspace{-0.2cm}

\begin{IEEEbiography}
[{\includegraphics[width=1in,height=1.25in,clip,keepaspectratio]{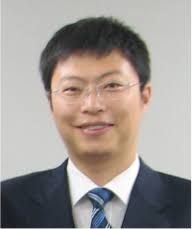}}]{Haiwei Dong}
(Senior Member, IEEE) received the Ph.D. degree from Kobe University, Kobe, Japan in 2010 and the M.Eng. degree from Shanghai Jiao Tong University, Shanghai, China, in 2008. He was a Principal Engineer with Artificial Intelligence Competency Center in Huawei Technologies Canada, Toronto, ON, Canada, a Research Scientist with the University of Ottawa, Ottawa, ON, Canada, a Postdoctoral Fellow with New York University, New York City, NY, USA, a Research Associate with the University of Toronto, Toronto, ON, Canada, and a Research Fellow (PD) with the Japan Society for the Promotion of Science, Tokyo, Japan. He is currently a Principal Researcher with Ottawa Research Center, Huawei Technologies Canada, Ottawa, ON, Canada, and a registered Professional Engineer in Ontario. His research interests include artificial intelligence, multimedia, metaverse, and robotics. He also serves as a Column Editor of IEEE Multimedia Magazine; an Associate Editor of ACM Transactions on Multimedia Computing, Communications, and Applications; and an Associate Editor of IEEE Consumer Electronics Magazine.
\end{IEEEbiography}
 \vspace{-1cm}

 \begin{IEEEbiography}
[{\includegraphics[width=1in,height=1.25in,clip,keepaspectratio]{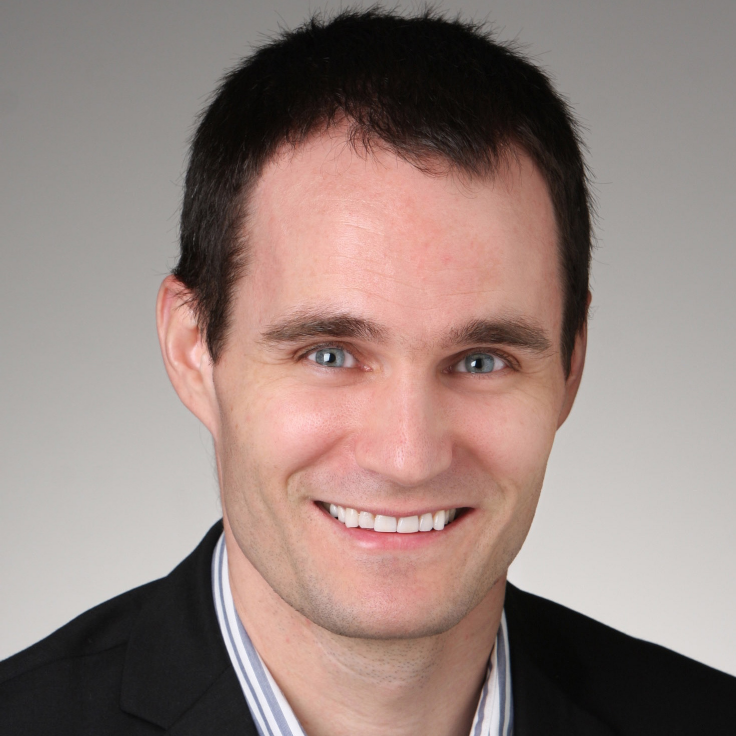}}]{Marc St-Hilaire}
(Senior Member, IEEE) received the Ph.D. degree in computer engineering from Polytechnique Montreal in 2006. He is currently a Professor with the School of Information Technology with a cross-appointment to the Department of Systems and Computer Engineering at Carleton University, Ottawa, Canada. He is conducting research on various aspects of wired and wireless communication systems. More precisely, he is interested in network planning and design, network architecture, mobile computing, and cloud/edge computing. With more than 170 publications, his work has been published/presented in several journals and international conferences. Over the years, he has received several awards, including the Carleton Faculty graduate mentoring award, the Carleton teaching achievement award, and several best paper awards. Finally, Dr. St-Hilaire is actively involved in the research community. In addition to serving as a member of technical program committees of various conferences, he is equally involved in the organization of several national and international conferences and workshops. He is also a senior member of the IEEE.
\end{IEEEbiography}
 \vspace{-1cm}
 
\begin{IEEEbiography}
[{\includegraphics[width=1in,height=1.25in,clip,keepaspectratio]{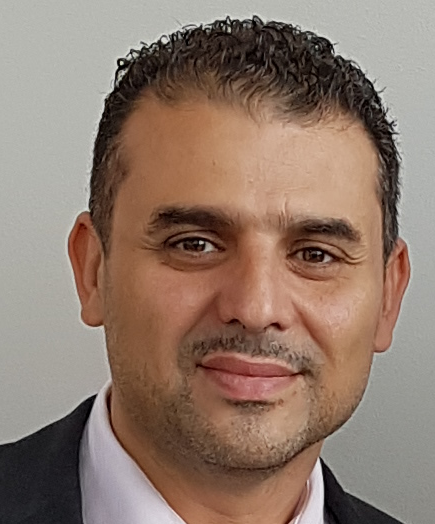}}]{Abdulmotaleb El Saddik}
(Fellow, IEEE) is currently a Distinguished Professor with the School of Electrical Engineering and Computer Science, University of Ottawa and a Professor with Mohamed bin Zayed University of Artificial Intelligence. He has supervised more than 120 researchers. He has coauthored ten books and more than 550 publications and chaired more than 50 conferences and workshops. His research interests include the establishment of digital twins to facilitate the well-being of citizens using AI, the IoT, AR/VR, and 5G to allow people to interact in real time with one another as well as with their smart digital representations. He received research grants and contracts totaling more than \$20 M. He is a Fellow of Royal Society of Canada, a Fellow of IEEE, an ACM Distinguished Scientist and a Fellow of the Engineering Institute of Canada and the Canadian Academy of Engineers. He received several international awards, such as the IEEE I\&M Technical Achievement Award, the IEEE Canada C.C. Gotlieb (Computer) Medal, and the A.G.L. McNaughton Gold Medal for important contributions to the field of computer engineering and science.
\end{IEEEbiography}
 \clearpage

\appendix


We provide the detailed protocol and Mean Opinion Score (MOS) results of the subjective user study in this appendix. The purpose of this study is twofold. First, it evaluates whether the proposed object-aware bitrate allocation is consistent with real user perception by comparing DIA to baseline approaches. Second, it examines the impact of the object update frequency by comparing DIA under different update frequencies.

The two subjective evaluations were conducted independently to answer different research questions. The baseline comparison evaluated DIA against existing approaches, whereas the frequency comparison evaluated different object update frequencies of DIA. Consequently, the reported MOS values were computed independently for each experiment and should be interpreted within their respective evaluation contexts. Minor differences in the absolute MOS values across the two experiments are therefore expected and do not reflect differences in the DIA configuration.

\subsection{Study Protocol}

We used processed gameplay clips from different game genres, including \textit{Cyberpunk 2077}, \textit{Spider-Man 2}, \textit{The Witcher 3}, \textit{Black Myth: Wukong}, and \textit{CS:GO}. Specifically, we selected two clips from \textit{Spider-Man 2} and \textit{Black Myth: Wukong}, and one clip from each of the remaining games. For the first group of evaluations, each gameplay clip was processed by DIA, Pensieve, T-Gaming, and MetaABR under the same video and network settings to ensure a fair subjective comparison. In this study, we used a bandwidth-constrained and time-varying network trace to emulate a congested access network. The available bandwidth for each chunk varied from 10 Mbps to 100 Mbps, with an average bandwidth of approximately 48 Mbps. The trace includes repeated low-bandwidth periods that trigger frequent quality adaptations, allowing us to examine whether object-aware bitrate allocation improves perceived quality under constrained network conditions. We used the same bandwidth trace for both DIA and baseline approaches, so that the MOS comparison reflects the effect of bitrate allocation rather than differences in network conditions. Different network conditions and bandwidth levels have been evaluated in the main simulation experiments; the subjective study focuses on user-perceived quality under a representative congested network setting.

Thirty-four participants were recruited to evaluate the perceived visual quality and overall viewing experience of the video clips through a web-based user study platform. For each input clip, participants compared the outputs generated by the evaluated approaches and assigned an individual MOS from 1 to 5 to each output clip, where a higher score indicates better perceived quality. The platform supports side-by-side viewing, focused A/B switching, and wipe-based comparison to help participants better perceive visual differences. To reduce ordering bias, the clips were presented in a randomized order for each participant. Each participant was required to complete ratings for all clips and all approaches before submitting the study, and the final MOS results were computed by averaging the completed participant ratings.


\subsection{MOS Comparison Among DIA and Baseline Approaches}

Table~\ref{tab:mos_baselines} summarizes the MOS results of DIA and three representative baseline approaches, including T-Gaming, Pensieve, and MetaABR. DIA consistently achieves the highest MOS across all evaluated gameplay clips. The average MOS values of DIA, Pensieve, T-Gaming, and MetaABR are 4.16, 2.49, 3.74, and 2.86, respectively.

Pensieve is included as a representative DRL-based ABR approach, MetaABR represents a recent learning-based ABR approach, and T-Gaming represents a content-aware cloud-gaming scheme. Pensieve and MetaABR perform bitrate adaptation at the frame level, while T-Gaming differentiates quality between salient ROIs and background regions. In contrast, DIA performs semantic object-level allocation according to user interests. The MOS results show that DIA achieves the highest perceived visual quality among the evaluated approaches.

To further examine the statistical significance of these improvements, we conducted paired t-tests using the paired MOS samples collected in the subjective study. The results of paired t-tests show that DIA significantly outperformed each baseline ($p<0.001$), demonstrating the effectiveness of the proposed object-aware bitrate allocation strategy.

\begin{table}[!ht]
\centering
\caption{MOS Comparison Among DIA and Baseline Approaches.}
\label{tab:mos_baselines}
\begin{tabular}{lcccc}
\toprule
Gameplay clip & DIA & Pensieve & T-Gaming & MetaABR\\
\midrule
\textit{Cyberpunk 2077} & 4.26 & 2.68 & 3.71 & 2.71\\
\textit{Spider-Man 2}-1 & 4.18 & 2.18 & 3.71 & 2.82\\
\textit{Spider-Man 2}-2 & 4.18 & 2.61 & 3.82 & 2.94\\
\textit{The Witcher 3} & 4.21 & 2.45 & 3.91 & 3.00\\
\textit{Black Myth: Wukong}-1 & 4.18 & 2.64 & 3.82 & 2.85\\
\textit{Black Myth: Wukong}-2 & 4.21 & 2.45 & 3.50 & 2.94\\
\textit{CS:GO} & 3.91 & 2.41 & 3.68 & 2.76\\
\midrule
Average & \textbf{4.16} & 2.49 & 3.74 & 2.86\\
\bottomrule
\end{tabular}
\end{table}

\subsection{Consistency Between the Proposed QoE Metrics and MOS}

\begin{table*}[!t]
\centering
\caption{QoE Results of Different ABR Approaches Across the Gameplay Clips Used in the Subjective Study. Each Entry Reports $\mathrm{QoE}_{\mathrm{lin}} /
 \mathrm{QoE}_{\log} /
 \mathrm{QoE}_{\mathrm{HD}}$.}
\label{tab:qoe_game_clips}
\resizebox{\textwidth}{!}{
\begin{tabular}{lcccc}
\hline
\textbf{Clip} & \textbf{DIA} & \textbf{Pensieve} &
\textbf{T-Gaming} & \textbf{MetaABR} \\
\hline
CS:GO
& \textbf{97.437 / 4.442 / 14.952}
& 11.463 / 2.446 / 0.961
& 34.045 / 3.424 / 11.952
& 22.288 / 2.626 / 11.955 \\

Cyberpunk 2077
& \textbf{53.112 / 3.832 / 13.239}
& 11.893 / 2.332 / 0.963
& 28.045 / 3.117 / 7.525
& 14.424 / 2.574 / 1.959 \\

Spider-Man 2-1
& \textbf{25.774 / 2.619 / 5.955}
& 10.346 / 2.150 / 0.960
& 8.056 / 2.001 / 1.227
& 5.482 / 2.164 / 1.013 \\

Spider-Man 2-2
& \textbf{98.156 / 4.434 / 14.736}
& 13.104 / 2.444 / 0.959
& 53.000 / 3.766 / 11.808
& 41.334 / 3.204 / 7.173 \\

The Witcher 3
& \textbf{57.589 / 3.926 / 13.544}
& 11.373 / 2.201 / 0.961
& 18.561 / 2.955 / 4.744
& 8.878 / 2.225 / 2.453 \\

Wukong 1
& \textbf{25.248 / 2.695 / 7.277}
& 12.509 / 2.382 / 0.959
& 7.303 / 2.390 / 4.120
& 8.092 / 1.302 / 5.139 \\

Wukong 2
& \textbf{51.042 / 3.591 / 11.573}
& 12.448 / 2.377 / 0.959
& 3.402 / 2.722 / 3.395
& 6.599 / 2.292 / 2.955 \\
\hline
\end{tabular}}
\end{table*}

\begin{table}[t]
\centering
\caption{Average QoE Metrics Across All Game Clips.}
\label{tab:average_qoe}
\begin{tabular}{lccc}
\hline
\textbf{Approach} &
$\mathbf{QoE}_{\mathrm{lin}}$ &
$\mathbf{QoE}_{\log}$ &
$\mathbf{QoE}_{\mathrm{HD}}$ \\
\hline
DIA
& \textbf{58.337}
& \textbf{3.648}
& \textbf{11.611} \\
Pensieve
& 11.877
& 2.333
& 0.960 \\
T-Gaming
& 21.773
& 2.911
& 6.396 \\
MetaABR
& 15.300
& 2.341
& 4.664 \\
\hline
\end{tabular}
\end{table}
To evaluate the consistency between the proposed QoE metrics and the subjective MOS values, we calculate the three QoE metrics for the four ABR approaches across all game clips under the same network settings in the subjective user study. The QoE results for each game clip are shown in Table~\ref{tab:qoe_game_clips}, while the average results are summarized in Table~\ref{tab:average_qoe}. As shown in these two tables, all three QoE metrics demonstrate a consistent performance trend across the evaluated ABR approaches: DIA achieves the highest average QoE, followed by T-Gaming, MetaABR, and Pensieve. This ranking is consistent with the average MOS results of 4.16, 3.74, 2.86, and 2.49, respectively. 

More specifically, while the overall ranking of the four approaches is consistent between the objective QoE metrics and MOS, some differences can be observed for the lower-performing approaches. In particular, Pensieve and MetaABR are not always ranked consistently by $\mathrm{QoE}{\mathrm{lin}}$ and $\mathrm{QoE}{\log}$ at the individual-clip level. One possible reason is that both approaches produce relatively low subjective quality, with average MOS values of 2.49 and 2.86, respectively. When the overall visual quality is already low, small differences in bitrate-related objective metrics may not be easily distinguishable by users. In contrast, $\mathrm{QoE}{\mathrm{HD}}$ shows better consistency with the MOS ordering. This suggests that a QoE formulation that gives greater emphasis to the quality of key objects may better reflect user-perceived visual quality than metrics dominated by bitrate magnitude alone.

\subsection{A Representative Example}

\begin{figure*}[!ht]
\centering
\includegraphics[width=0.78\linewidth]{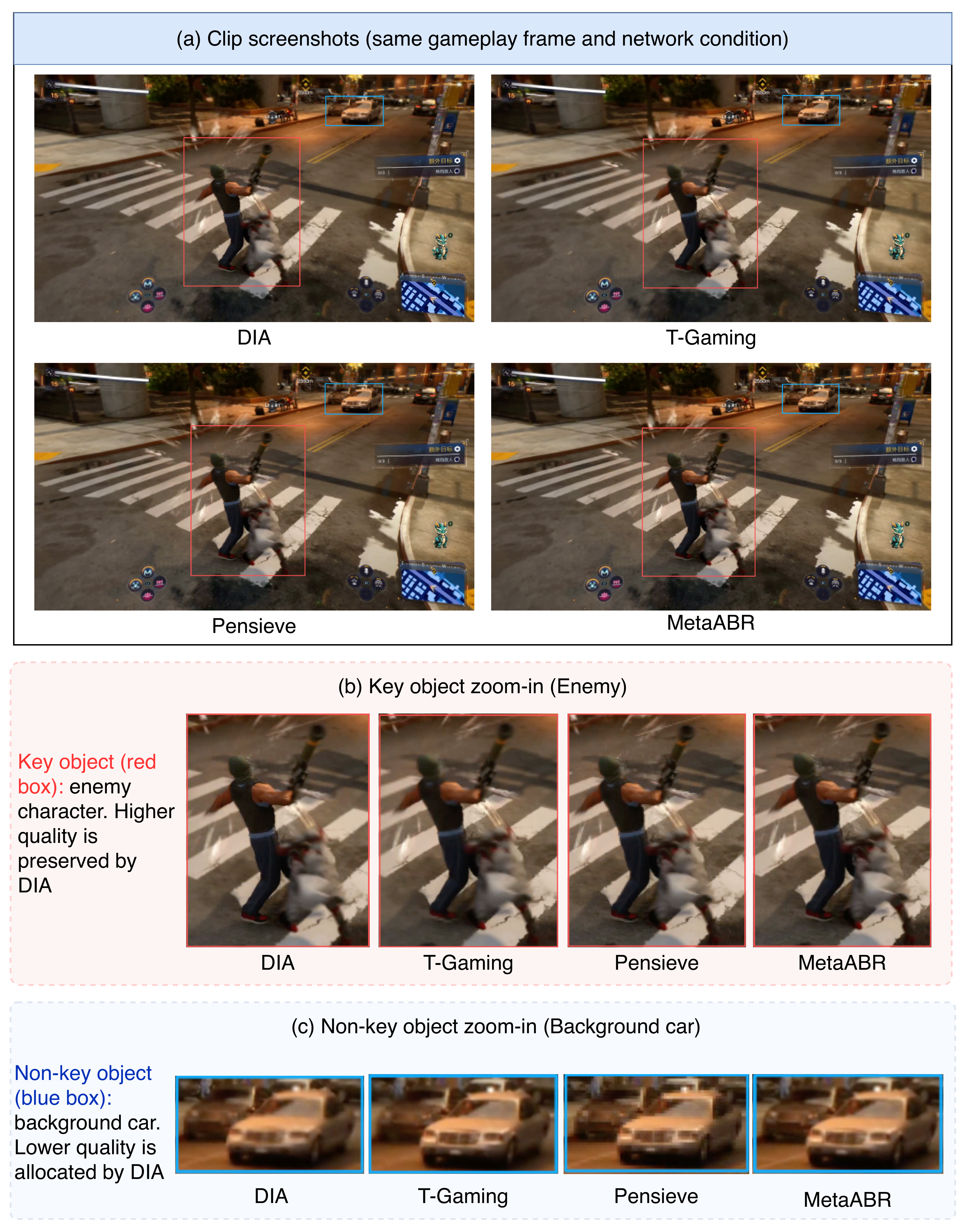}
\caption{Visual comparison among DIA and three representative baseline approaches under the same gameplay frame and network condition. Panel (a) presents the full-frame results. The red bounding box highlights a key object, namely the enemy character in this example, while the blue bounding box highlights a non-key object (background car). Panel (b) enlarges the key object, and Panel (c) enlarges the non-key object. Pensieve and MetaABR perform frame-level bitrate adaptation and therefore apply a uniform quality level to the entire frame. T-Gaming performs ROI-level quality differentiation based on visual saliency. In contrast, DIA performs semantic object-level bitrate allocation, obtaining higher quality for key objects while assigning lower quality to non-key objects.}
\label{fig:visual_comparison}
\end{figure*}

Fig.~\ref{fig:visual_comparison} presents a representative qualitative comparison among DIA, Pensieve, T-Gaming, and MetaABR under the same gameplay frame and network condition. Pensieve and MetaABR perform bitrate adaptation at the frame level, where a single quality level is uniformly applied to the entire frame. Consequently, they cannot explicitly differentiate between gameplay-relevant objects and less important background content.

T-Gaming introduces spatial quality differentiation at the ROI level by preserving visually salient regions at a higher quality than the background. However, its bitrate allocation is driven primarily by visual saliency rather than the semantic and gameplay relevance of individual objects. Therefore, it cannot explicitly assign different quality levels to objects according to their roles in the gameplay.

In contrast, DIA performs semantic object-aware bitrate allocation. As shown in Fig.~\ref{fig:visual_comparison}(b), DIA preserves higher visual quality for the gameplay-relevant object highlighted by the red bounding box, namely the enemy character in this example. Meanwhile, Fig.~\ref{fig:visual_comparison}(c) shows that DIA assigns a lower quality to the less important background vehicle highlighted by the blue bounding box. By reducing the bitrate allocated to less important objects, DIA reserves more bandwidth for objects that are more relevant to user attention and gameplay experience under constrained network conditions. This qualitative observation is consistent with the subjective MOS results presented in Table~\ref{tab:mos_baselines}.

\subsection{MOS Comparison Under Different Object Update Frequencies}

Table~\ref{tab:mos_frequency} summarizes the MOS results under different object update frequencies. The average MOS values of DIA under 2~Hz, 4~Hz, and 8~Hz are 3.70, 4.05, and 4.17, respectively. A Friedman test confirms that the update frequency has a statistically significant impact on the subjective scores ($p<0.001$). Pairwise comparisons further show that both 4~Hz and 8~Hz significantly outperform 2~Hz ($p<0.001$), while no statistically significant difference is observed between 4~Hz and 8~Hz ($p=0.096$).

Although 8~Hz achieves the highest MOS, its improvement over 4~Hz is relatively small (0.12 MOS on average) and is not statistically significant. Therefore, 4~Hz is selected as a practical default setting, as it provides a good balance between object-level responsiveness and computational overhead.

\begin{table}[!ht]
\centering
\caption{MOS Comparison Under Different Object Update Frequencies.}
\label{tab:mos_frequency}
\begin{tabular}{lccc}
\toprule
Gameplay clip & 2~Hz & 4~Hz & 8~Hz \\
\midrule
\textit{Cyberpunk 2077} & 3.33 & 3.83 & 4.25 \\
\textit{Spider-Man 2}-1 & 4.08 & 4.25 & 4.17 \\
\textit{Spider-Man 2}-2 & 3.83 & 4.00 & 4.08 \\
\textit{The Witcher 3} & 3.25 & 3.75 & 4.00 \\
\textit{Black Myth: Wukong}-1 & 4.08 & 4.25 & 4.25 \\
\textit{Black Myth: Wukong}-2 & 3.92 & 4.42 & 4.58 \\
\textit{CS:GO} & 3.42 & 3.83 & 3.83 \\
\midrule
Average & 3.70 & 4.05 & 4.17 \\
\bottomrule
\end{tabular}
\end{table}

\subsection{Summary}

The subjective user study provides two main findings. First, DIA achieves higher MOS values than all baseline approaches across the evaluated gameplay clips, which supports the perceptual benefit of prioritizing key objects. Second, increasing the object update frequency from 4~Hz to 8~Hz only brings limited MOS improvement. Therefore, 4~Hz is adopted as a practical setting that balances perceptual quality and computational overhead.

\end{document}